\documentclass[10pt]{iopart}
\usepackage{iopams}
\usepackage[utf8]{inputenc}
\usepackage{color}
\usepackage{overpic}
\usepackage{siunitx}
\usepackage{soul}
\usepackage[dvipsnames]{xcolor}
\usepackage{iopams}

\usepackage{graphicx}
\usepackage[numbers]{natbib}

\usepackage{float}
\usepackage{graphicx}
\graphicspath{{./figures/}}
\usepackage[export]{adjustbox}
 \usepackage{booktabs}
 
\usepackage{pgfplots}
\pgfplotsset{compat=1.15}
\usepackage{pgfplotstable} 
\usepgfplotslibrary{external} 
\usepackage[breaklinks=true]{hyperref}
\hypersetup{
  unicode=false,
  pdftoolbar=true,
  pdfmenubar=true,
  pdffitwindow=false,
  pdfstartview={FitH},
  pdfsubject={Runaway electron generation in ITER mitigated disruption with improved physics models},
  pdfcreator={Lorenzo Votta et al},
  pdfproducer={Lorenzo Votta et al},
  pdfkeywords={Tokamak} {Disruptions} {Runaway electrons} {DREAM} {Shattered pellet injection},
  pdfnewwindow=true,
  colorlinks=true,
  linkcolor=blue,
  citecolor=blue,
  filecolor=blue,
  urlcolor=blue
}

\newcommand{\DREAM}{\textsc{Dream}}

\newcommand{\JOREK}{\textsc{Jorek}}
\newcommand{\ASTRA}{\textsc{Astra}}
\newcommand{\CORSICA}{\textsc{Corsica}}

\newcommand{\eqref}[1]{(\ref{#1})}

\begin{document}

\title[Runaway electrons in ITER]{Runaway electron generation in ITER mitigated disruptions with improved physics models}

\author{L. Votta$^{1}$, F. J. Artola$^{3}$, E. Nardon$^{4}$, O. Vallhagen$^{2}$, M. Hoppe$^{1}$ }

\address{$^{1}$Department of Electrical Engineering, KTH Royal Institute of Technology, SE-100 44 Stockholm, Sweden}
\address{$^{2}$ Department of Physics, Chalmers University of Technology,
SE-41296 Gothenburg, Sweden}
\address{$^{3}$ ITER Organization, Route de Vinon-sur-Verdon, CS 90 046, 13067 St. Paul Lez Durance Cedex, France}
\address{$^{4}$CEA, IRFM, F-13108 Saint-Paul-lez-Durance, France}
\ead{votta@kth.se}
\vspace{10pt}

\begin{abstract}
We assess runaway-electron (RE) generation in ITER disruptions mitigated by shattered pellet injection (SPI) using improved physics modelling in the 1D disruption simulation framework \DREAM. To this end, we extend \DREAM\ with four ITER-relevant physics models: (i) a reduced model for RE scrape-off associated with the vertical plasma motion, (ii) a semi-analytical plasmoid-drift model for material deposition, (iii) an adaptive hyper-resistive transport model to suppress unphysical thin-current channels during the current quench (CQ), and (iv) an updated Compton RE generation seed calculated for the new ITER tungsten first-wall design. We simulate full-current \qty{15}{MA} L-mode (H26, non-nuclear) and H-mode (DTHmode24, nuclear) scenarios, and an intermediate-current \qty{7.5}{MA} H-mode non-nuclear case, from realistic ITER inputs. Complete avoidance of a multi-MA RE beam is found to require a long pre-thermal quench (TQ) duration to thermalize the hot-tail electrons, high deuterium assimilation with limited neon, and a representative seed current comparable to a single RE in ITER. As previously found with lower fidelity setups [Vallhagen et al, Nucl. Fusion 64
(2024)], these conditions are met by staggered or low-Ne injections in H26, but are typically violated in DT H-mode when nuclear seeds are present. In addition to analyzing the effect of the new models, we investigate the role of the current spike  associated with the TQ and importance of radial transport of runaways in the CQ. After incorporating these additional physical effects into a comprehensive disruption model and analyzing their impact, we present a representative ITER DT H-mode SPI scenario which provides a theoretically viable route to tolerable RE currents in ITER fusion power operation.

\end{abstract}

\ioptwocol

\section{Introduction}

Tokamak disruptions represent one of the greatest physics and engineering challenges for the reliable operation of ITER and future fusion reactors. During these off-normal events, the stored thermal and magnetic energy of the plasma is rapidly released, while a strong toroidal electric field is induced, potentially generating multi-megampere beams of relativistic runaway electrons (REs). Therefore, a robust disruption mitigation system (DMS) is an essential component for machine protection. To be effective, the system must minimize localized heat loads on the vessel wall during the thermal quench (TQ), protect it from runaway electron (RE) beam impacts, and limit the electromagnetic forces exerted on structural components. In a 15 MA ITER plasma, these objectives translate into specific operational constraints: the current quench (CQ) must complete within 50–150 ms to limit structural forces, and the RE current must be suppressed below \qty{150}{kA} to ensure that localized energy deposition remains within material tolerance~\cite{Lehnen2021}. This limit is, however, scenario-dependent: it depends on the assumed RE termination/impact characteristics, in particular the toroidal extent (axisymmetry vs.\ localized strikes) of the deposition footprint and on whether a significant fraction of the plasma magnetic energy is converted into additional RE kinetic energy during the terminal loss~\cite{Pitts2025}.

The disruption mitigation strategy of ITER relies on shattered pellet injection (SPI) \cite{Lehnen2021}. This technique, first demonstrated at DIII-D in 2010 \cite{Commaux2010}, involves the rapid injection of massive quantities of cryogenic material---specifically, a mixture of neon and hydrogen for ITER---as soon as a disruption is expected to occur. The injected material cools the plasma and increases collisional drag, before significant RE populations can develop. In recent years, much work has gone into developing SPI as a disruption mitigation technique \cite{Commaux2016,Baylor2019,Combs2018,Meitner2017} and at present several key tokamaks operate, or have operated, their own SPI systems, including DIII-D \cite{Meitner2017}, ASDEX Upgrade \cite{Dibon2023}, KSTAR \cite{Park2020}, and JET \cite{Wilson2020,Herfindal2019}, which provide data to guide the design of the ITER DMS \cite{Herfindal2019,Papp2020,Jachmich2021}. However, a critical challenge lies in predicting how SPI performance will scale from current experimental conditions to reactor-scale plasma parameters. While existing experiments provide insights into material assimilation and plasma response, the nonlinear physics governing RE dynamics, particle and heat transport, and MHD stability under ITER-relevant conditions introduces significant uncertainties that cannot be resolved through empirical extrapolation alone.

Predictive modeling becomes essential to bridge this gap between present experimental capabilities and reactor-scale requirements. Several studies have performed integrated simulations of complete disruption events, focusing on the CQ phase and commonly relying on simplified geometries and/or prescribed profiles for material deposition and temperature reduction~\cite{martin-solis_loarte_lehnen_2017,Vallhagen2020, pusztai_ekmark_bergstrom_halldestam_jansson_hoppe_vallhagen_fulop_2023}. Early results~\cite{martin-solis_loarte_lehnen_2017} indicated that sufficiently large hydrogen injections could suppress RE production. However, later work~\cite{Vallhagen2020} showed that at very high hydrogen densities the accompanying recombination reduces the free-electron density. This enhances the avalanche gain by offsetting the competition between frictional drag and avalanche multiplication~\cite{Hesslow2019b}, leading to the production of multi-megaampere RE currents.

A first investigation into SPI-mitigated disruptions in ITER using a self-consistent treatment of the impurity injection and TQ, linked to the evolution of magnetic perturbations, was presented in~\cite{Vallhagen2022}. This study aimed to assess the impact of dividing the injection into two separate phases, building on earlier work in Ref.~\cite{Nardon2020}. The resulting two-stage cooling effectively thermalized the high-energy tail of the electron distribution before the onset of RE generation, thereby significantly reducing the runaway seed population from the hot tail mechanism. This approach showed potential for achieving low RE currents in scenarios where activated seed mechanisms such as tritium beta decay and photon Compton scattering from activated wall materials are absent. It should be noted that these simulations were limited to a small subset of scenarios and incorporated several idealizations, including circular plasma cross sections, a fixed timing for the TQ independent of plasma evolution, and the omission of ion transport following pellet ablation. Recent simulations incorporating enhanced transport coefficients informed by 3D modeling have shown that, in 15 MA non-nuclear ITER scenarios, both single and staggered SPI can effectively suppress the runaway current below critical thresholds~\cite{Vallhagen2024}. However, for \qty{15}{MA} deuterium-tritium (DT) plasmas, the runaway current consistently exceeds the mega-ampere level when employing the same mitigation strategies.

While these results represent significant progress in understanding RE mitigation, they remain affected by key modeling limitations, which have motivated the present study. First, earlier simulations neglected vertical plasma motion, despite its ubiquity during disruptions in ITER~\cite{Artola2024}. However, recent \JOREK\ simulations~\cite{Wang2024, Bandaru2025} have demonstrated that vertical plasma motion during the CQ can induce significant RE losses through scrape-off, as magnetic flux surfaces intersect the vessel wall. To capture this effect, a scrape-off loss model was introduced in \DREAM\ in~\cite{Vallhagen2025}, and validated against two-dimensional magnetohydrodynamic (2D-MHD) simulations with \JOREK. 
 
Second, the modelling of material assimilation following SPI in earlier works neglected the expected drift of ablated material in the curved magnetic geometry of the tokamak \cite{Pgouri2006,Lang1997,Parks2000,Vallhagen2023}. This likely resulted in an overestimation of material deposition in the plasma core and, consequently, an overprediction of mitigation efficiency. A semi-analytical model capturing this drift was developed in~\cite{Vallhagen2023}, implemented in \DREAM, and subsequently validated against ASDEX Upgrade experiments~\cite{Vallhagen2025b}.
 
Third, the one-dimensional fluid models employed in \DREAM\ are prone to the formation of unphysically narrow current channels, which can artificially extend the TQ via localized ohmic heating~\cite{Putvinski1997}. This one-dimensional artifact is expected to be suppressed by inherently three-dimensional MHD dynamics. In the present work, we introduce a model in \DREAM\ that mimics the effect of such MHD instabilities.
 
Finally, the Compton scattering RE seed model employed in previous works was based on outdated beryllium (Be) first wall configurations and required revision in light of the new ITER tungsten (W) first-wall design. 

In this work, we present a computationally efficient modeling framework that simultaneously addresses these four effects, while also accounting for the multitude of physics found to be relevant in previous works. Section~\ref{sec:physicsmodels} describes the integration of physics-based models for vertical motion-induced scrape-off, plasmoid drift, MHD-driven enhanced transport, and an updated Compton seed source into the 1D disruption simulation framework \DREAM~\cite{Hoppe2021}, along with the general disruption model and the ITER scenarios simulated in this study. In Section~\ref{sec:results}, we evaluate the combined impact of the implemented models on the performance of SPI disruption mitigation in ITER. Finally, in Section~\ref{sec:discussion} we conclude by extending the analysis to a representative nuclear ITER disruption scenario, highlighting conditions under which RE suppression may be achieved.

\label{sec:introduction}

\section{Improved physics models}
\label{sec:physicsmodels}

To simulate disruptions in ITER, a baseline model is set up using \DREAM. The model incorporates a wide range of physical effects, including those used in Ref.~\cite{Vallhagen2024} as well as a number of new effects which have been proven to be of great importance in ITER. In this section, we give an overview of the disruption model used and describe the physics of the four main new physics models implemented for the purpose of this study~\cite{Vallhagen2024}.

\begin{figure*}
    \begin{overpic}[width=\linewidth]{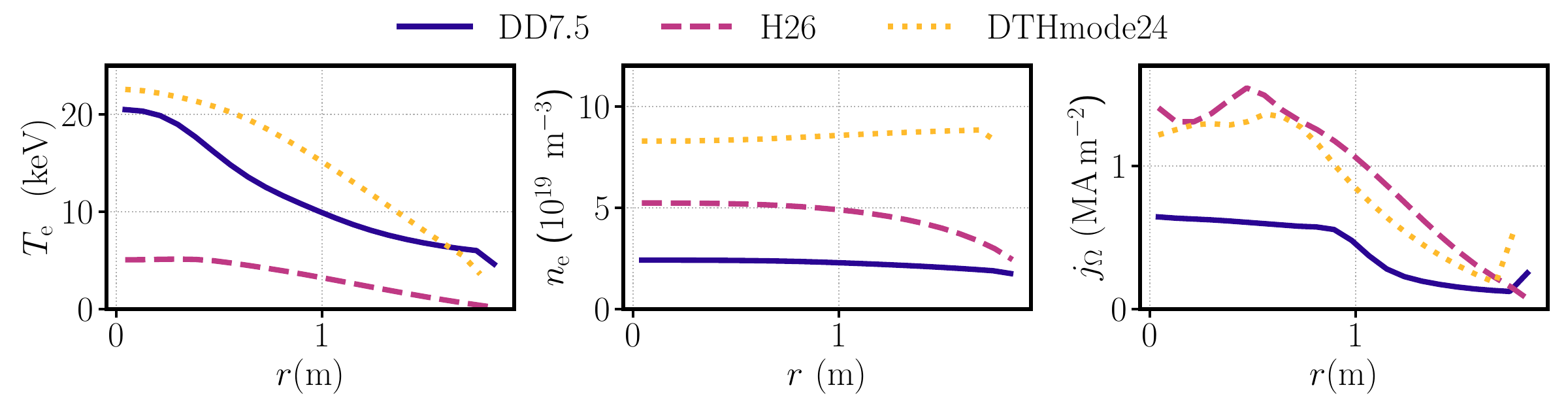}
        \put(29,19){(a)}
        \put(62,19){(b)}
        \put(95.3,19){(c)}
    \end{overpic}
\caption{Initial plasma profiles for the scenarios considered. Solid blue lines correspond to the intermediate current scenario (\emph{DD7.5}), dashed magenta lines to \emph{H26}, and dotted yellow lines to \emph{DTHmode24}. Shown are (a) electron temperature, (b) electron density, and (c) ohmic current density as functions of the radial coordinate.}
\label{fig:initialProf}
\end{figure*}

The simulations are based on three ITER scenarios described in Refs.~\cite{Lehnen2013,IDM28WPTG}. Two of these are \qty{15}{MA} scenarios, referred to as full current cases: \emph{DTHmode24}, a deuterium-tritium H-mode plasma with equal isotope concentrations, and \emph{H26}, a hydrogen L-mode plasma that, unlike \emph{DTHmode24}, does not include nuclear RE seeds, as no nuclear reactions are expected. Both these scenarios feature a \qty{15}{MA} plasma current and will hence be referred to as the \emph{full current} scenarios. The third scenario, referred to as the \emph{intermediate current} scenario, is \emph{DD7.5MA}, a pure deuterium H-mode plasma, featuring a plasma current of \qty{7.5}{MA}. In all simulations, numerical magnetic equilibria are used with a wall radius of $b = \qty{2.833}{m}$, chosen to reproduce the poloidal magnetic energy inside the vacuum vessel from \JOREK\ simulations, including corrections for poloidal field coil contributions. The wall is assumed to be resistive with a wall time of $\tau_{\rm wall}=\qty{0.5}{s}$. Initial plasma profiles are imported from the transport code \CORSICA~\cite{Kim_2018} for the full current scenarios and from \ASTRA~\cite{Fable2013} for the intermediate current scenario, and are shown in figure~\ref{fig:initialProf}.

\subsection{Disruption model}\label{sec:theory:disruption}
\paragraph{Shattered pellet injection}
The SPI model implemented in \DREAM\ is described in detail in Ref.~\cite{Vallhagen2022}. The mean pellet fragment velocity is set to $v_{\rm p} = 500 \rm m/s$, with a uniform velocity spread of $\Delta v/v_{\rm p} = 0.4$, and an injection divergence angle of $\alpha = \pm 10^{\circ}$. Fragments are assumed to originate from $(R, Z) = (8.568, 0.6855)\, \rm {m}$. We analyze a single ITER pellet containing $1.85 \times 10^{24}$ atoms using only the shard size corresponding to 487 fragments. The fragment size distribution follows the formulation of Parks~\cite{Parks2016}. In all scenarios, the number of atoms per shard is adjusted to conserve the total pellet mass. After pellet shattering, the ablation of individual shards is modeled using the Neutral Gas Shielding (NGS) approach~\cite{parks2017}. The ablation rate includes a correction for the empirical magnetic-field reduction factor of the form $(2/B)^{0.843}$ (with $B$ in tesla), which originates from cylindrically symmetric near-field FronTier simulations of pellet ablation in magnetized
plasmas that neglect plasmoid drift effects~\cite{Bosviel2020FronTier,ZhangParks2020ArealDeposition}. Since 3D simulations including $\nabla B$-driven drift show a weaker magnetic-field reduction of the ablation rate than axially symmetric models, applying this fit to ITER conditions may underestimate the ablation rate when drift effects are important. Deposition locations are further modified by plasmoid drift, modeled according to~\cite{Vallhagen2023} and briefly described in section ~\ref{sec:plasmoidDrift}. Besides cooling the plasma through radiation and dilution by the deposited material, SPI also tends to destabilize MHD modes, potentially triggering a TQ.

\paragraph{Thermal quench}
In this work we use the term TQ to mean the phase during which cross-field transport is strongly enhanced, leading to the rapid loss of thermal energy and increased transport of REs. The “TQ onset” refers to the time at which the enhanced-transport is activated. Two distinct criteria are used to trigger the TQ in the simulations. With the \emph{early} onset criterion, the TQ begins when a neon-carrying shard reaches the flux surface with $q=2$. With the \emph{late} onset criterion, the TQ is instead initiated when the temperature at any location within the $q=2$ surface falls to $10 \, \rm eV$ or below. These numbers are motivated by the observation that resistive current profile decay becomes sufficiently rapid at $\qty{10}{eV}$ to potentially trigger MHD instabilities near the $q=2$ surface within approximately one millisecond. Notably, the local temperature of the plasmoid or at given field line can drop to $\qty{10}{eV}$ before the flux-surface-averaged value does, particularly near Ne shards traversing the $q\approx2$ region, which may induce MHD activity, captured by the early onset condition. Accordingly, we regard the \emph{early} onset condition as the most conservative trigger, allowing a prompt TQ from localized cooling, whereas the \emph{late} onset condition corresponds to the more optimistic limit in terms of RE generation.

Once the TQ starts,  spatially and temporally constant electron heat transport and runaway electron diffusion are activated for the TQ duration $t_{\rm TQ}$, set to either $t_{\mathrm{TQ}} = \qty{1}{ms}$ or \qty{3}{ms}, in line with experimental extrapolations for ITER scenarios. We refer to this phase as the TQ or TQ transport phase. These transport processes are modeled using Rechester-Rosenbluth-like diffusion coefficients dependent on the normalized magnetic perturbation amplitude, $\delta B/B$~\cite{RRtransport}. The $\delta B/B$ values are chosen to ensure that the central electron temperature decreases below $\qty{200}{eV}$ within the TQ duration, assuming transport as the only energy loss channel. In practice, additional losses such as radiation further reduce the temperature. At sufficiently low temperatures ($ \approx 100 \, \rm eV$), ohmic heating balances transport losses, preventing further cooling due to transport alone, and further cooling is instead dominated by radiation. The $\delta B/B$ values used, listed in Table~\ref{tab:dBB}, typically range between 0.1\% and 1\%, consistent with 3D non-linear MHD simulations under ITER-like conditions~\cite{Hu2021SPI_Radasym,Särkimäki2020}.

Ion transport for all charge states is also initiated concurrently with the heat and runaway electron transport. The initial ion diffusion coefficient is $D_{\mathrm{ion}} = \qty{4000}{m^2/s}$, and the advection coefficient is $A_{\mathrm{ion}} = -\qty{2000}{m/s}$. These coefficients then decay exponentially to zero with a characteristic timescale $\tau = \qty{0.5}{ms}$. The negative sign of \(A_{\rm ion}\) corresponds to an inward pinch which, together with the large initial \(D_{\rm ion}\), drives a rapid core accumulation of neon and thus a radially peaked neon density profile which may not be fully representative of expected neon profiles. Similar transport models have been validated against ASDEX Upgrade experimental data using \ASTRA\ simulations. The characteristic transport times $\tau_{D}=a^2 / D_{\mathrm{ion}}$ and $\tau_{A}=a / A_{\mathrm{ion}}$, with $a$ the plasma minor radius, also lie in the millisecond range. This configuration supports rapid core material redistribution within $\sim\qty{0.1}{ms}$, consistent with 3D MHD modeling~\cite{Nardon2020,Hu2021SPI_Radasym}.

\paragraph{Runaway electrons}
The \DREAM\ simulations presented here employ a fully fluid model for RE dynamics. Dreicer generation is calculated using a neural network trained on kinetic simulations~\cite{Hesslow2019}, while hot-tail generation is calculated according to Appendix C of Ref.~\cite{Hoppe2021}. Avalanche multiplication is implemented according to the model in~\cite{Hesslow2019NF}, incorporating partial screening effects.

Runaway electron radial transport is modeled using the approach of Svensson~\cite{Svensson}, which averages velocity-resolved transport coefficients over momentum space for application to the fluid RE population. The momentum dependence is taken as $p_\parallel/(1 + p^2)$, capturing the expected low-energy $\propto v$ scaling and mimicking the suppression of transport at high energies due to finite orbit width effects~\cite{Särkimäki2020}. 
In the \emph{DTHmode24} scenario, additional RE generation sources from tritium beta decay and Compton scattering are included. To model the sharp drop in photon flux following cessation of fusion neutron generation, the Compton seed term is treated with a two-stage flux model: the photon flux is initialized using the ITER nominal value from~\cite{Martin-Solis2017}, and reduced by a factor of $10^{-4}$ at the end of the TQ, coinciding with the termination of the transport phase. This reduction mimics the expected decline in wall photon emission when neutron bombardment ceases, while accounting for the delayed photon flux.

Finally, to capture RE losses during the vertical plasma motion, the reduced scrape-off loss model from~\cite{Vallhagen2025} is included, along with an additional term in the electron heat equation to account for parallel heat losses in open field line regions, as described in section~\ref{sec:theory:vde}.

\subsection{Plasmoid drift model}
\label{sec:plasmoidDrift}
As a pellet ablates in the hot plasma, its cold, dense ablation cloud (plasmoid) drifts across magnetic field lines towards the low-field side due to charge separation in the curved tokamak field, which generates a vertical electric field driving the plasmoid outward. This drift can carry material away from the core, especially for pure hydrogenic pellets whose relatively hot (tens of eV), and thus highly over-pressurized, plasmoids experience strong outward motion. In contrast, neon-doped pellets form cooler clouds with minimal drift, resulting in deposition closer to the ablation point.

The plasmoid drift model used here was developed in~\cite{Vallhagen2023}. In that work, the radial displacement of the ablation cloud $\Delta r$ is expressed as an integral of the electric field component $E_y$ perpendicular to the cloud surface:
\begin{equation}
    \Delta r = \frac{1}{B} \int_0^\infty E_y(t)\,\mathrm{d}t,
    \label{eq:D1.2:Deltar}
\end{equation}
where $B$ is the magnetic field strength. We obtain $E_y(t)$ by integrating the linear evolution equations given in Eqs. (2.31)–(2.32) of Ref.~\cite{Vallhagen2023}. Specifically, in \DREAM\ we evaluate $\Delta r$ in the analytical limit of negligible Alfvén currents expressed in Eq.~(A4) of Ref.~\cite{Vallhagen2023}.

This model assumes that the ablation cloud is  a slab with uniform poloidal thickness that drifts and expands along magnetic field lines. The plasma properties are assumed to be uniform across both the poloidal and toroidal directions. To define the state of the drifting material, several parameters must be specified: the initial temperature $T_0$ near the pellet, a characteristic drift temperature $T$, the mean ionization state $\langle Z_i\rangle$ for each ion species $i$ present in the pellet composition, and the half-width $\Delta y$ of the drifting cloud. These quantities can be inferred from existing experimental data \cite{Muller} and computational studies \cite{Matsuyama_2022, Vallhagen2025b}. For the scenarios considered in this work, we adopt $T_0 = 2 \, \rm eV$, $\langle Z_{\rm H} \rangle =1$, $\langle Z_{\rm Ne} \rangle=2$, with a drift temperature $T=5 \, \rm eV$ for neon-doped pellets and $T=30 \, \rm eV$ for pellets composed purely of protium. The choice of $T_0$ reflects the threshold at which the ablation cloud becomes sufficiently ionized to initiate drift. The lower drift temperature in the Ne-doped case arises from enhanced radiative cooling around \qty{5}{eV}, which suppresses further heating. Nevertheless, this temperature is sufficient to nearly fully ionize hydrogenic species, justifying the selected value of $\langle Z_{\rm H}\rangle =1$. Among the parameters involved, the half-width $\Delta y$ has the most significant influence on the drift behavior of the ablation cloud. Although precise measurements are challenging, both experimental observations and numerical simulations consistently suggest that $\Delta y \approx 12.5 \, \rm mm$ is a reasonable and representative estimate \cite{Muller,Parks1996,Kong2024,Samulyak2021,Matsuyama_2022, Vallhagen2025b}.

\subsection{Modelling of Vertical Displacement Events}
\label{sec:theory:vde}
To model vertical displacement events in \DREAM, we have implemented two complementary models. The first model accounts for RE losses due to the vertical displacement of the plasma, while the second model addresses parallel heat losses in the halo region when the flux surfaces are open.

\subsubsection{Scrape-off runaway electron loss model}
The scrape-off RE loss model incorporated into our simulations aims to capture the essential effects of flux surface scrape-off during vertical displacement events. The model builds upon observations from \JOREK\ simulations, which show that the variation in the poloidal magnetic flux at the instantaneous last closed flux surface (LCFS) during the CQ in ITER is relatively minor~\cite{Wang2024}. This consistency in poloidal flux allows a simplified modeling framework that distinguishes between open and closed field line regions, even though the plasma geometry for a given toroidal flux surface is fixed. The model was originally derived, implemented, and tested in~\cite{Vallhagen2025}.
To enforce the condition that REs in the open field line region are lost rapidly, a particle loss term is introduced for the RE density ($n_\mathrm{RE}$) on a flux surface labeled by $r$:
\begin{equation}\label{eq:scrapeoff}
    \left( \frac{\partial n_\mathrm{RE}}{\partial t} \right)_{\text{scrape-off}} = - \frac{n_\mathrm{RE}}{\tau_\text{loss}} \Theta(r - r_\mathrm{LCFS}),
\end{equation}
where $\tau_\text{loss}$ represents the characteristic loss time scale, $\Theta$ is the Heaviside step function, and $r_\mathrm{LCFS}$ is the flux surface label corresponding to the LCFS. The flux surface $r_\mathrm{LCFS}$ is determined by solving $\psi_{\mathrm{p}}(r_\mathrm{LCFS}) = \psi_{\mathrm{p}}(a, t = 0)$, where $a$ is the initial minor radius of the plasma, and $\psi_\mathrm{p}(a, t = 0)$ denotes the poloidal flux at the edge of the plasma at the beginning of the disruption. A physically motivated choice for the loss time scale is $\tau_\text{loss} \sim R_0/c$, where $R_0$ is the major radius of the tokamak and $c$ is the speed of light. However, this time scale is much shorter than any other relevant time scale during the disruption, except for short intervals during rapid ionization or recombination. To avoid increasing computational requirements unnecessarily, the more practical choice $\tau_\text{loss} \sim 10 \Delta t$ is employed, where $\Delta t$ is the numerical time step used in the simulations. This choice maintains computational tractability without altering the qualitative behavior of the scrape-off process on the time scales ($t \gtrsim
 t_\mathrm{TQ}$). 
\subsubsection{Parallel heat loss model}

Disruptions with low radiation levels can cause significant heat loads during vertical displacements in the CQ phase, as seen in JET with its ITER-like wall~\cite{Lehnen2013}. Experimental observations have shown that such heat fluxes are related to the halo currents generated in the region of open field lines~\cite{Pautasso1994}. The DINA code~\cite{Lukash1996} was previously updated to include convective power losses in the halo region, in addition to radiation losses, by incorporating parallel heat fluxes of the form $q_{\parallel} \sim n_e T_e c_{\mathrm{s}}$~\cite{Kiramov2016}, and this has now also been implemented in \DREAM\ as a sink term in the electron energy balance equation, consistent with the EU-DEMO current-quench heat-flux model of Ref.~\cite{Pautasso_2025}:
\begin{equation}
   \left( \frac{\partial W_\mathrm{cold}}{\partial t} \right)_{\rm halo}= - \frac{4}{3}  \frac{\gamma_{\mathrm{sh}} n_e c_{\mathrm{s}} T_e}{L_\parallel} \Theta(r-r_\mathrm{LCFS}),
\end{equation}
where $W_\mathrm{cold}$ represents the electron thermal energy density, $c_{\mathrm{s}} = \sqrt{(T_e + \gamma T_i)/m_i}$ is the sound speed and $\gamma_{\mathrm{sh}} = 8$ is the heat transmission coefficient used in the simulations. The ion mass used to calculate the speed of sound is approximated to the majority species mass, while $\gamma = 5/3$ for an adiabatic transformation with isotropic pressure, and the term $L_\parallel = 2 \pi q R_0$ is the length of the flux tube along the magnetic field. \DREAM\ calculations including this heat loss term have been verified against the results in~\cite{Artola2024}, showing a relative error of $\lesssim 3\%$ in the radiation-to-ohmic power ratio.

\subsection{Adaptive hyper-resistivity model}\label{sec:theory:hypres}
A major challenge in 1D CQ simulations is the formation of narrow spikes in the ohmic current density, which produce short, unphysical bursts of ohmic heating that re-heat the plasma and prevent a complete TQ. The intense local Ohmic heating raises the electron temperature, reducing the plasma resistivity and further concentrating the current into an increasingly narrow channel. One such artifact is the formation of thin, radially localized high--current--density channels, as noted by Putvinski~\cite{Putvinski1997}.
The result is unphysical plasma re-heating, preventing simulation completion and yielding inaccurate results. In real 3D tokamak plasmas, such sharp current gradients are MHD unstable and rapidly lead to enhanced radial transport, thereby flattening gradients and suppressing the instability. Since \DREAM\ does not inherently capture these 3D MHD effects, an adaptive transport model has been implemented to emulate this essential mechanism.

The model continuously monitors the radial profile of the total current density, $j_{\rm tot}$. If the gradient of this profile exceeds a prescribed threshold, the enhanced transport is triggered. The trigger condition is based on the normalized current density gradient:
\begin{equation}
    \left|\nabla j_\mathrm{tot}^\mathrm{norm}\right| = \left|\frac{a}{\langle j_{\rm tot}\rangle} \frac{\partial j_{\rm tot}}{\partial r}\right|,
\end{equation}
where $a$ denotes the plasma minor radius and $\langle j_{\rm tot} \rangle$ is the radially averaged total current density. Once exceeded, the model triggers a global enhanced transport for heat, runaway density \footnote{Neglected in actual simulations to be conservative on runaway generation.}, and poloidal magnetic flux. Rather than simply adding an arbitrary diffusion coefficient to fix the numerical issue, the model links all transport effects to a single physical parameter: the normalized magnetic perturbation strength, $\delta B/B$, which represents the level of magnetic stochasticity that would be generated by the MHD instability. For the electron thermal energy density $(W_{\rm cold} = 3/2 n_{\rm cold}T_{\rm cold})$ and RE density $(n_{\rm re})$, a diffusive transport term is applied using the Rechester-Rosenbluth \cite{RRtransport} diffusion coefficient
\begin{equation}
\label{eq:rechesterRosenbluth}
		D_{\rm re} = \pi qR_{\rm m}v_\parallel\left(\frac{\delta B}{B}\right)^2,
\end{equation}
for transport in stochastic fields, where $q$ is the safety factor, $R_{\rm m}$ is the major radius and $v_\parallel$ is the electron parallel velocity equal to the speed of light in vacuum $c$ for RE and to the thermal speed $v_\mathrm{th}$ for bulk electrons. Simultaneously, to represent the effect of field line breaking and reconnection, a hyper-resistive diffusion is applied to the poloidal flux, $\psi_{\rm p}$. The hyper-resistivity, $\Lambda_{\rm m}$, is not arbitrary but is calculated to be physically consistent with the chosen $\delta B /B$, based on the formula \cite{Boozer2018} 
\begin{equation}\label{eq:LambdaHypres}
		\Lambda =
			\frac{1}{144}
			\frac{\mu_0}{4\pi}
			\frac{V_{\rm A}\psi_{\rm t}^2}{N_{\rm t}}
        				        		        =
        \frac{\pi \mu_0 qR_{\rm m}^2 V_{\rm A}\psi_{\rm t}^2}{288a^2}
			\left(
				\frac{\delta B}{B}
			\right)^2.
\end{equation}
where $V_{\rm A}=B/\sqrt{\mu_0\rho}$ is the Alfvén velocity,
$\psi_{\rm t}$ is the toroidal flux enclosed by the plasma, and $N_{\rm t}$ the number of toroidal turns needed for a stochastic field line to travel across the plasma. We note that a stochastization of the magnetic field would likely lead to transport of REs, however to remain conservative in our estimate of the number of surviving REs, we choose to neglect this transport when activating the hyper-resistivity model to suppress narrow current density channels.

\subsection{Updated Compton scattering photon spectra}

\begin{figure*}
    \centering
    \begin{overpic}[width=\textwidth]{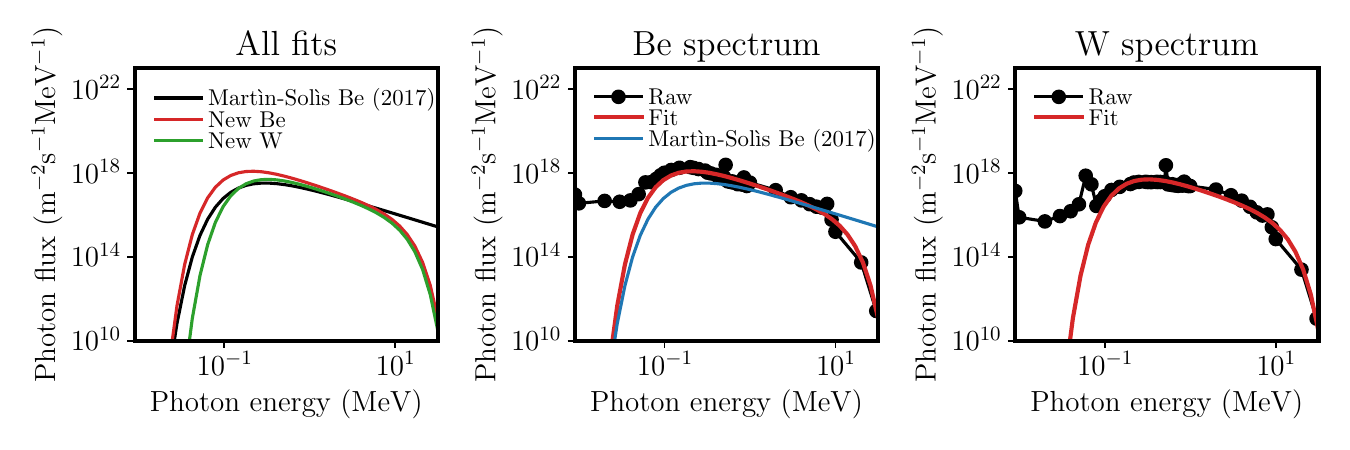}
        \put(9.8,29.4){(a)}
        \put(42.4,29.4){(b)}
        \put(75.0,29.4){(c)}
    \end{overpic}
    \caption{Fitted Compton energy spectra for different ITER first wall materials. (a) In black the original spectrum from \cite{Martin-Solis2017}, in red the new Be spectrum, and  in green the new W spectrum. (b) In black, the raw data from the new radiation transport simulations for Be, in red the new fit to this data, and in blue the old fit by Martìn-Solìs {\em et al}. (c) In black, the raw data from the new radiation transport simulations for W, and in red the new analytical fit to this spectrum.}
    \label{fig:ComptonSpectra}
\end{figure*}
In high-current ITER disruptions, a significant source of runaway seeds is expected to be Compton scattering, where high-energy gamma rays, produced from neutron interactions with the first wall, scatter off thermal electrons and accelerate them to runaway energies. Previous \DREAM\ simulations \cite{Vallhagen2021,Vallhagen2024,Ekmark2024,Vallhagen2025,Vallhagen2025b} used the gamma photon spectrum given by Martín-Solìs {\em et al}~\cite{Martin-Solis2017} to evaluate the runaway generation via Compton scattering, which was provided for a Be wall. However, the reference design for the ITER first wall has since been updated to consist entirely of W. As the gamma spectrum and flux are highly dependent on the interacting material~\cite{Reali2023}, this change in the baseline design necessitates a corresponding update to the Compton seed model in \DREAM. 

In the present work, three photon spectra are used as input, corresponding to prompt gamma emission for a nominal 500~MW ITER fusion power plasma. Specifically, we consider the original Be spectrum from Martín-Solís \emph{et al.}~\cite{Martin-Solis2017}, a more recent calculation for a Be first wall, and a newly generated spectrum for a W first wall~\cite{ComptonData}. The fitted Compton $\gamma$-ray energy spectra are shown in Fig.~\ref{fig:ComptonSpectra}.
  Photons with energies below \qty{0.1}{MeV} are excluded from the fits, as their contribution to RE generation is negligible. In the intermediate energy range, the updated Be photon spectrum generally exhibits higher flux than the previously used spectrum, indicating that the old Be data underestimated the Compton $\gamma$-ray flux in this domain. In contrast, the W spectrum shows closer agreement with the old Be spectrum, featuring a slightly higher flux in the mid-energy range and a reduced flux at lower energies. Due to these relatively small deviations, the impact on Compton runaway seed generation is expected to be minimal. In this work, the new W spectrum has been used for the simulations.

Furthermore, the photon flux decreases by a factor of $F_{\gamma} = 10^{4}$ shortly after fusion reactions terminate, corresponding to the cessation of neutron bombardment at the wall and the end of the TQ transport phase in \DREAM\ simulations. To account for this temporal variation, the simulations are initialized with the nominal ITER photon flux reported in~\cite{Martin-Solis2017}, which is then reduced by a factor of $10^{4}$ at the conclusion of the transport event. In earlier simulations, a similar approach was employed, but with a reduction factor of $10^{3}$. Notably, this choice does not substantially affect the resulting RE current in our simulations.

\section{Simulation results}
\label{sec:results}

This section investigates the influence of the new physical models introduced in Section~\ref{sec:physicsmodels} on RE suppression in realistic ITER scenarios during SPI. Building on the parameter scan of Ref.~\cite{Vallhagen2024} (summarized in Table~\ref{tab:scenarios}), we extend the analysis with additional targeted studies. Specifically, after having repeated the parameter scan of Ref.~\cite{Vallhagen2024} utilizing new physics models described above, we focus on four key aspects: (i) the sensitivity of RE suppression to reduced neon injection content, (ii) the role of hyper-resistive diffusion during the TQ, (iii) the influence of RE transport throughout the CQ, and (iv) the RE mitigation efficiency at reduced plasma current in ITER.

%%%%%%%%%%%%%%%%%%%%%%%%%%%%%%%%%%%%%%%%%%%%%%%%%%%%%%%%%%%%%%%%%%%%%%%%%%%%%%%%%%%%%%%%%%%%%%%%%%%%%%%%%%%%%%%%%%%%%%%%%%%%%%%%%%%%%%%%%%%%%%%%%%%%%%%%%%%%%%%%%%%%%%%%%%%%%%%%%%%%%%%%%%%%%%%%%%%%%%%%%%%%%%%%%%%%%%%%%%%%%%%%%%%%%%%%%%%%%%%%%%%%%%%%%%%%%%%%%%%%%%%%%%%%%%%%%%%%%%%%%%%%%%
\subsection{Impact of new models }
Here we apply the models described in section~\ref{sec:physicsmodels} to the full current (\qty{15}{MA}) DTHmode26 and H26 scenarios. Simulations with single,  multiple and staggered SPI schemes are considered, using up to four pellets delivered either in a single stage or two staggered stages. In the staggered scheme, pellets are injected sequentially with a fixed \qty{5}{ms} delay. In the multiple-injection scheme, several pellets are injected simultaneously from the same location, while keeping the total injected neon quantity constant (redistributed among the pellets). We label these schemes as S, St, and M: S denotes a single Ne doped pellet, St a two-step staggered injection (protium followed by a Ne doped pellet), and M simultaneous injection of multiple Ne doped pellets at the beginning of the pre-TQ phase. The index (e.g.\ 1-3) identifies the corresponding variants in Table~\ref{tab:scenarios} (pellet number/partitioning, Ne/H mixture, and assumed TQ conditions).

We evaluate four figures of merit to compare runaway mitigation performance. The \emph{pre-TQ time} is defined as the time difference between the transport event trigger and the time at which the first pellet shard reaches the last closed flux surface. Second, the runaway seed current is assessed via the \emph{representative RE seed}, defined as the RE current remaining after the TQ plus any additional seed generated during the CQ. Third, the \emph{representative RE current} is taken at the point when runaways carry $95\%$ of the total current (or corresponding to the maximum RE current if the plateau is not reached). Finally, \emph{material assimilation} quantifies the fraction of neon and protium atoms assimilated by the plasma.

Repeating the baseline scan of Ref.~\cite{Vallhagen2024} for \qty{15}{MA} ITER plasmas with updated physics models reveals that complete avoidance of a multi-megaampere runaway beam is only possible within a confined region of parameter space and only in scenarios in which nuclear RE seeds are absent.

In particular, runaway suppression in our simulations is achieved only when two process-level conditions are simultaneously satisfied: (i) the pre-TQ duration must be long enough for the hot-tail seed to thermalize, and (ii) a sufficient assimilation of $N_{\mathrm{H,assim}}\gtrsim 3\times 10^{23}$ atoms must be achieved while limiting the neon inventory to $N_{\mathrm{Ne,assim}}\lesssim 10^{22}$ atoms, so that the electron density is increased without triggering a premature radiative collapse and without excessively increasing the number of target thermal electrons for avalanche multiplication. 
The extent to which conditions (i) and (ii) must be fulfilled is then set by the effective seed entering the CQ, leading to the condition (iii) the representative seed current must be sufficiently small, ideally comparable to the current carried by a single relativistic electron in ITER, implying no meaningful seed for avalanche multiplication. 
Although conditions (i) and (ii) are not fully independent and their correlation depends on the injection scheme, framing (iii) as an outcome-level requirement makes explicit that measures improving one of (i)–(ii) do not necessarily improve overall performance if they simultaneously deteriorate the other. Injection schemes that satisfy (i) and (ii) well enough to reach condition (iii) consistently achieve full RE suppression in our simulations when nuclear seeds are neglected.

The clearest examples are the L-mode H26 staggered injection cases (St1–St3) and the low-neon single-injection case S3. In contrast, scenarios that violate any one of the above thresholds tend to develop multi-megaampere runaway beams. For example, neon-rich single-pellet injections or simultaneous multi-pellet injections that lead to a short pre-TQ phase (or assimilate too much neon) invariably lead to substantial post-disruption RE current. A notable exception is the DTHmode24 multiple-injection cases without nuclear RE seeds: M7 and M8 suppress REs despite a short pre-TQ phase ($t_{\mathrm{pre\text{-}TQ}}\sim 1\,\mathrm{ms}$). After the transport event, the plasma remains sufficiently hot that the electric field stays sub-critical ($E<E_{\mathrm c}$). The subsequent pellet ablation therefore occurs in a still hot background and efficiently thermalizes the hot-tail seed before the final cooling stage, when $T\simeq 20\,\mathrm{eV}$ and $E$ can transiently exceed $E_{\mathrm c}$.

\begin{figure*}
    \centering
    
    \begin{overpic}[width=\textwidth]{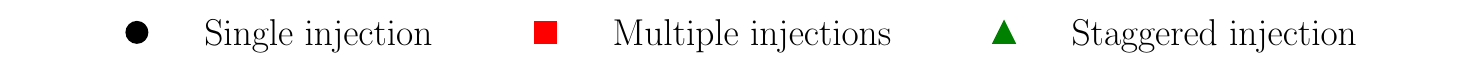}
    
    \end{overpic}\\
    \textbf{L-mode H}
    \vspace{3mm}

    \begin{overpic}[width=\textwidth]{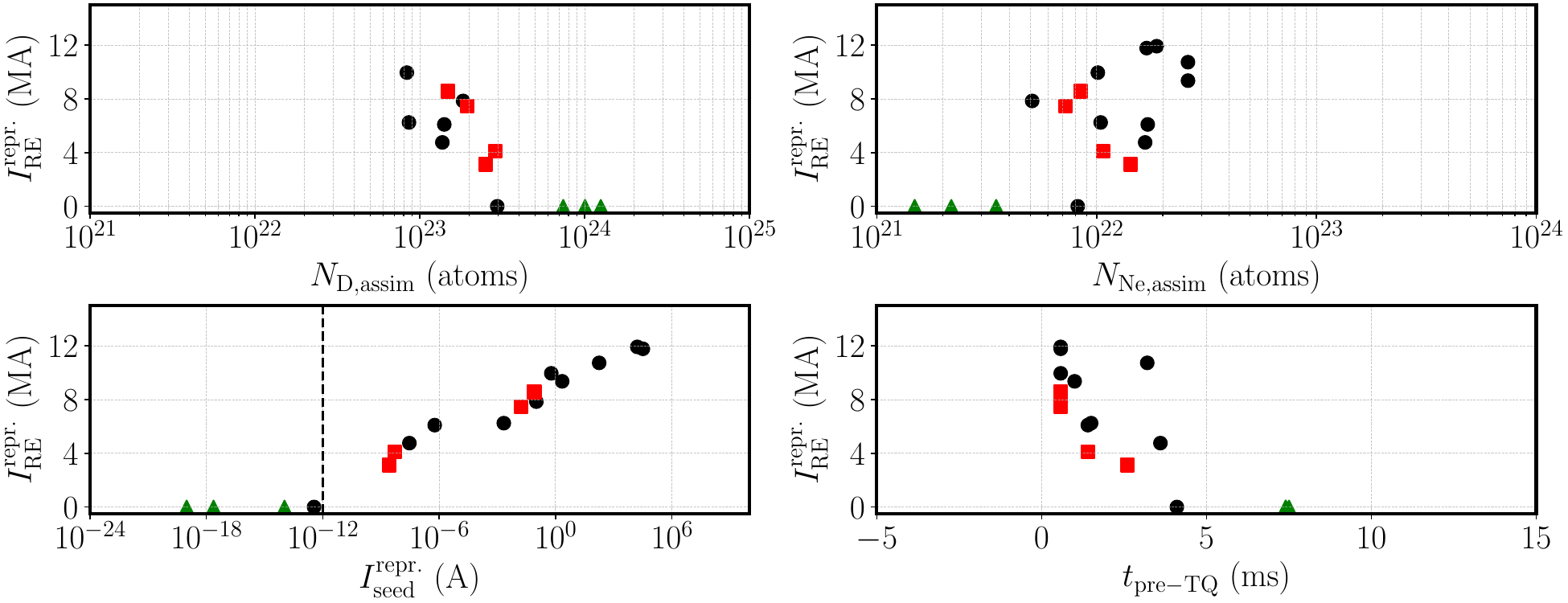}
                            \put(44,36){\textbf{(a)}}
        \put(94,36){\textbf{(b)}}
    \put(44,17){\textbf{(c)}}
    \put(94,17){\textbf{(d)}}
        \put(21,17){one electron}
    \put(21,15){in ITER}
    \end{overpic}
    \caption{
Correlations between the representative runaway current, $I_{\mathrm{RE}}^{\mathrm{repr}}$, and key quantities in the baseline SPI simulations for the L-mode H26 scenario. Panel (a) shows $I_{\mathrm{RE}}^{\mathrm{repr}}$ versus the number of assimilated protium atoms, $N_{\mathrm{H,assim}}$, while panel (b) shows $I_{\mathrm{RE}}^{\mathrm{repr}}$ versus the number of assimilated neon atoms, $N_{\mathrm{Ne,assim}}$. Panel (c) presents $I_{\mathrm{RE}}^{\mathrm{repr}}$ as a function of the representative runaway seed current, $I_{\mathrm{seed}}^{\mathrm{repr}}$, and panel (d) shows it versus the pre-TQ duration $t_{\mathrm{pre\text{-}TQ}}$. Each point represents a simulation and colours distinguishing injection schemes are indicated in the top legend.}
\label{fig:correlationsH26}

\end{figure*}
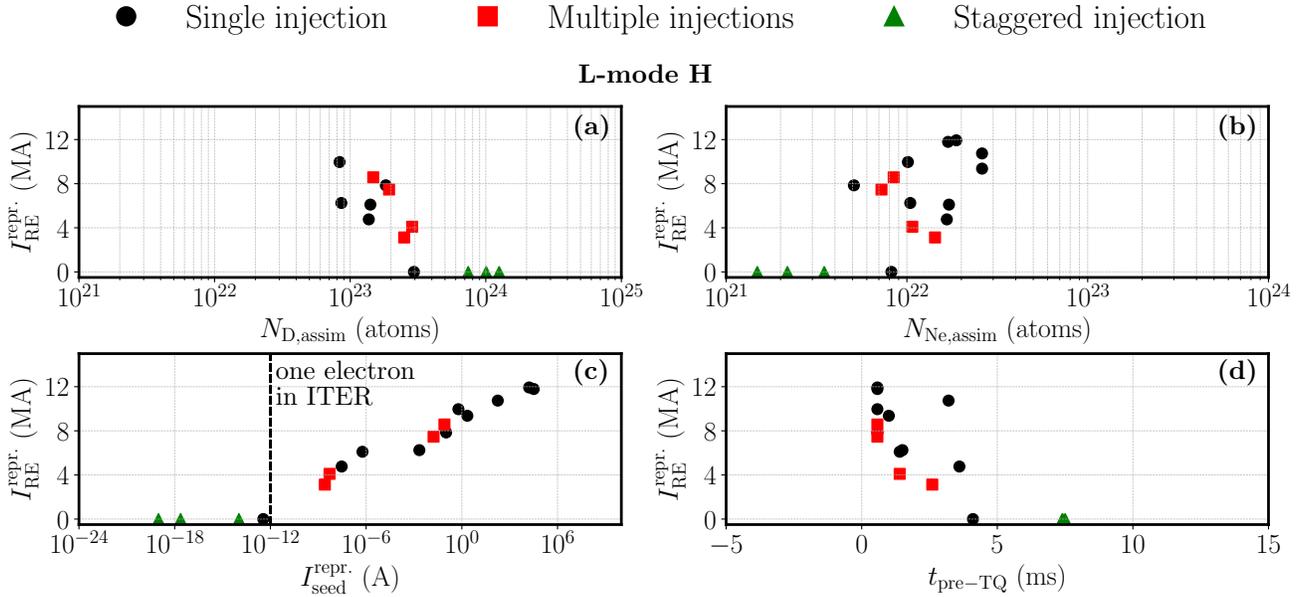
\subsubsection{Correlations among the explored parameters}\label{sec:correlations}
\begin{figure*}
    \begin{overpic}[width=\textwidth]{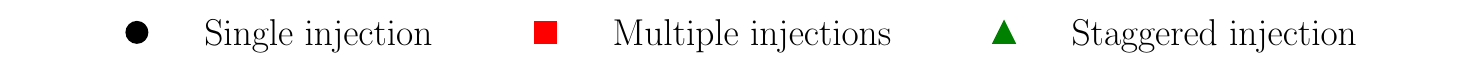}
    \end{overpic}\\
    \centering
    \textbf{H-mode D-T}
    \vspace{3mm}
    
    \begin{overpic}[width=\textwidth]{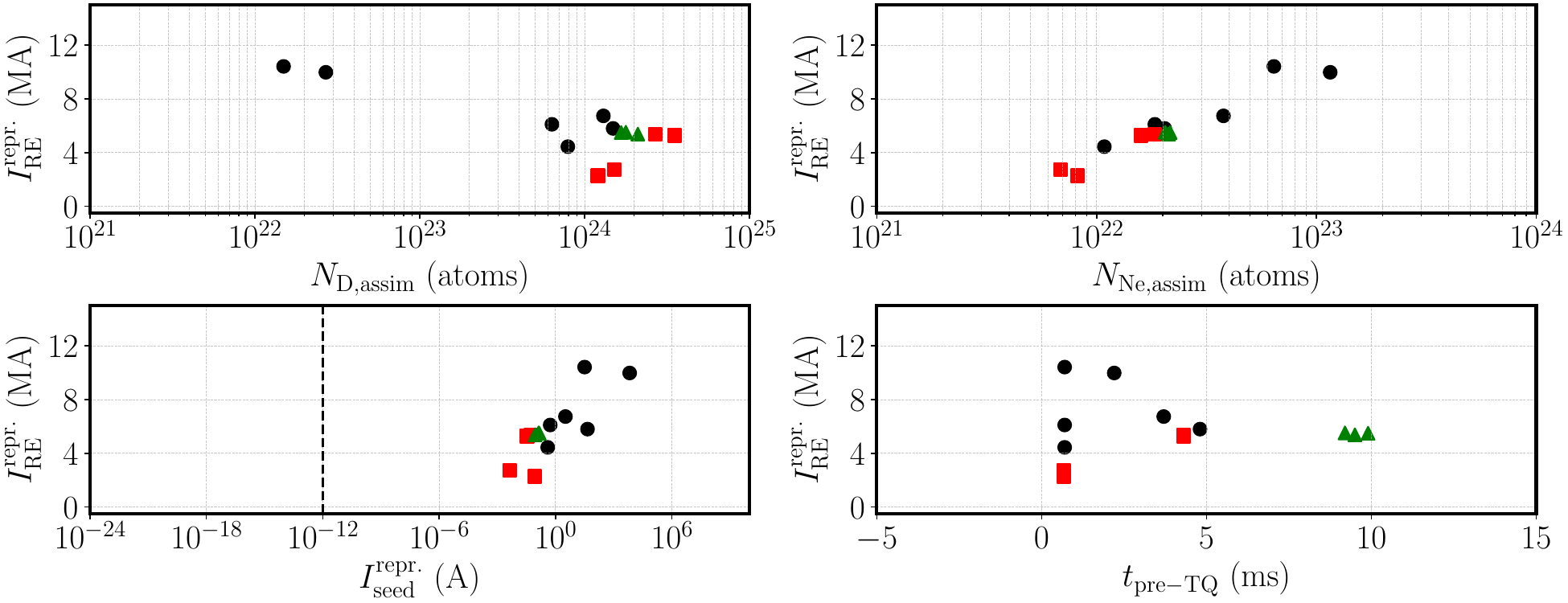}
    \put(44,36){\textbf{(a)}}
        \put(94,36){\textbf{(b)}}
    \put(44,17){\textbf{(c)}}
    \put(94,17){\textbf{(d)}}
        \put(21,17){one electron}
    \put(21,15){in ITER}
    \end{overpic}
    \caption{
        Correlations between the representative final runaway-electron current, $I_{\mathrm{RE}}^{\mathrm{repr}}$, and key quantities in the baseline SPI simulations for H-mode DTHmode24 plasmas. Panel (a) shows $I_{\mathrm{RE}}^{\mathrm{repr}}$ versus the number of assimilated protium atoms, $N_{\mathrm{H,assim}}$, panel (b) shows $I_{\mathrm{RE}}^{\mathrm{repr}}$ versus the number of assimilated neon atoms, $N_{\mathrm{Ne,assim}}$. Panels (c,d) present $I_{\mathrm{RE}}^{\mathrm{repr}}$ as a function of the representative seed current, $I_{\mathrm{seed}}^{\mathrm{repr}}$ and the pre-TQ duration $t_{\mathrm{pre\text{-}TQ}}$ respectively. Each point represents a simulation and colours distinguishing injection schemes are indicated in the top legend. 
    }
    \label{fig:correlationsDTH}
\end{figure*}

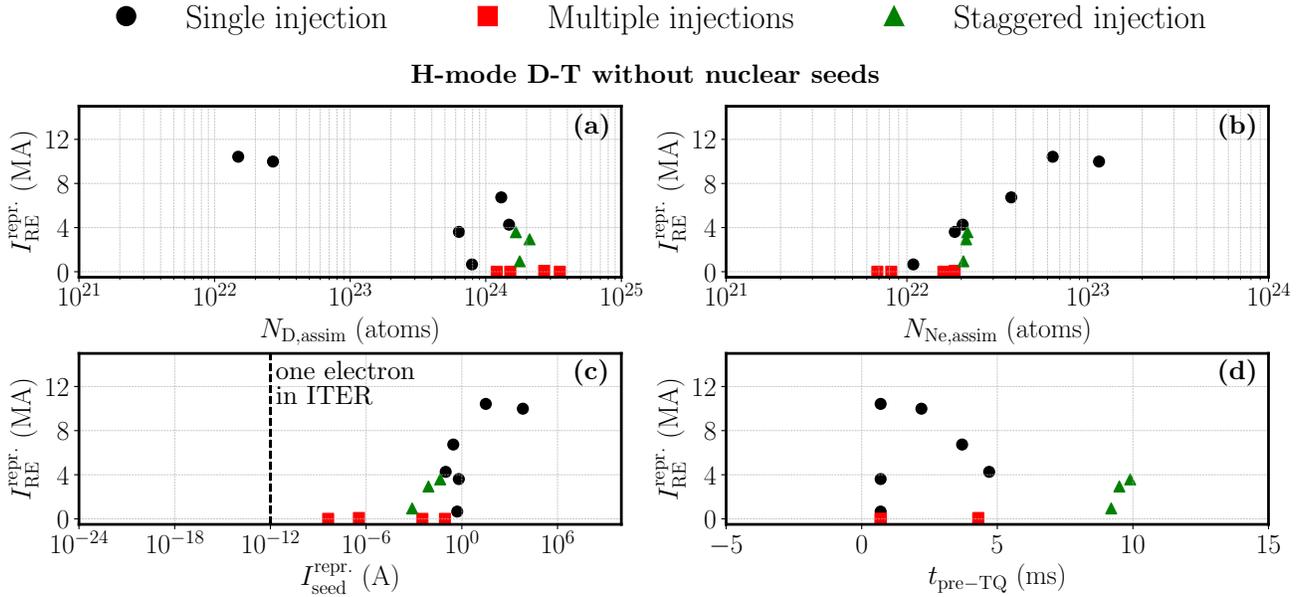
\begin{figure*}
    \begin{overpic}[width=\textwidth]{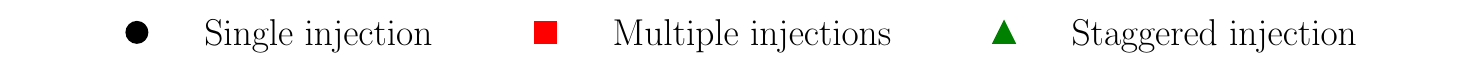}
    \end{overpic}
    \centering
    \textbf{H-mode D-T without nuclear seeds}
    \vspace{3mm}\\
    \begin{overpic}[width=\textwidth]{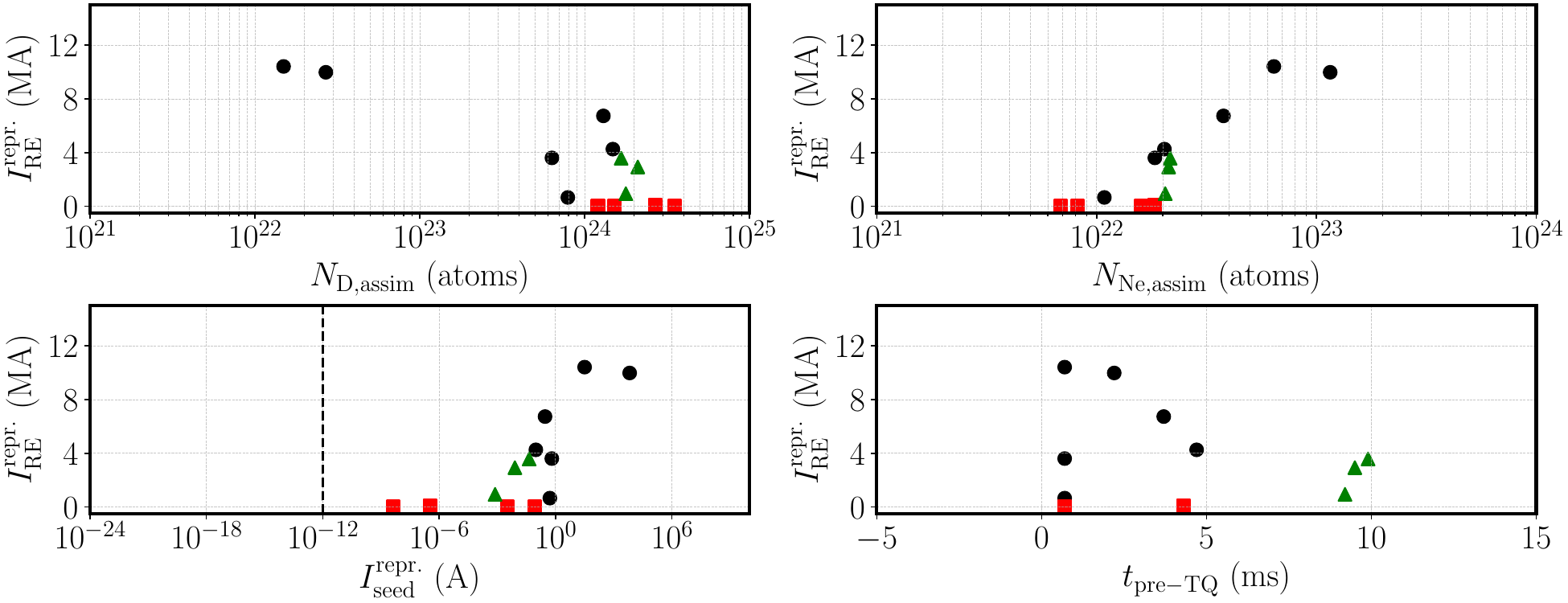}
                            \put(44,36){\textbf{(a)}}
        \put(94,36){\textbf{(b)}}
    \put(44,17){\textbf{(c)}}
    \put(94,17){\textbf{(d)}}
        \put(21,17){one electron}
    \put(21,15){in ITER}
    \end{overpic}
    \caption{
Correlations between the representative final runaway-electron current, $I_{\mathrm{RE}}^{\mathrm{repr}}$, and key quantities in the baseline SPI simulations for H-mode DTHmode24 plasmas without  nuclear seeds are shown. Panel (a) shows $I_{\mathrm{RE}}^{\mathrm{repr}}$ versus the number of assimilated protium atoms, $N_{\mathrm{H,assim}}$, panel (b) shows $I_{\mathrm{RE}}^{\mathrm{repr}}$ versus the number of assimilated neon atoms, $N_{\mathrm{Ne,assim}}$. Panels (c,d) present $I_{\mathrm{RE}}^{\mathrm{repr}}$ as a function of the representative seed current, $I_{\mathrm{seed}}^{\mathrm{repr}}$ and the pre-TQ duration $t_{\mathrm{pre\text{-}TQ}}$ respectively. Each point represents a simulation and colours distinguishing injection schemes are indicated in the top legend.
}

\label{fig:correlationsDTH-noact}
\end{figure*}

Figures~\ref{fig:correlationsH26} and~\ref{fig:correlationsDTH} show correlations between the representative runaway current, $I_{\text{RE}}$, and key figures of merit in the \qty{15}{MA} scenarios: the assimilated material content, the pre-TQ duration, and the seed current size. The correlation analysis confirms the finding from~\cite{Vallhagen2024} that the final runaway current depends on the generated runaway seed, shown in panel (c) of both figures, but the dependence is weak: across the explored cases, $I_\mathrm{RE}^\mathrm{repr.}$ varies only by an order-unity factor, while $I_\mathrm{seed}^\mathrm{repr.}$ spans $\sim 10 $ orders of magnitude. Nevertheless, since the achievable suppression of avalanche multiplication is bounded by the recombination limit of the assimilated material, successful RE mitigation still requires suppressing the seed to extremely low levels. Consequently, cases which minimize the seed are the most successful in preventing runaway formation. In the H26 scenario of figure~\ref{fig:correlationsH26}, this is achieved by all staggered injection schemes as well as the S3 single injection case, while in the DTHmode24 scenario no injection scheme suppresses the runaway current to below \qty{2}{MA}. When activated seeds are neglected in the DTHmode24 scenario---shown in figure~\ref{fig:correlationsDTH-noact}---all multiple pellet injection cases suppress the runaways, as do the S15 single injection and St5 staggered injection cases.

Single pellet injections generally result in significant post-disruption runaway currents. The reason for this is that the presence of neon in the pellet yields a short pre-TQ and a large runaway seed. The one exception is the H26 case S3, in which a number of favourable conditions interact to produce negligible runaway seed and representative runaway currents. In this case, a single pellet with a low neon concentration ($2.8\%$) is injected and ablates, depositing a significant amount of hydrogen before sufficient neon has been deposited to trigger a radiative collapse. As a consequence, the pre-TQ is extended beyond \qty{4}{ms}, which is approximately the time it takes for the pellet to reach the plasma core, giving ample time for most hot electrons to thermalize. In all simulations considered in this study, this condition on the pre-TQ time seems to be a necessary but not sufficient condition for all cases which successfully suppress runaway electrons. The subsequent TQ is set to last for \qty{3}{ms}, the longer of the two TQ durations considered among all cases, and thus provides sufficient time for any remaining seed runaway electrons to be expelled from the plasma before the CQ commences.

In the H26 L-mode scenario, simulations do not show an unambiguous overall advantage of multiple injections over single injections when considering $I_{\rm RE}^{\rm repr.}$ alone. A clearer trend emerges when considering the assimilated deuterium content: at fixed $N_{\mathrm{D,assim}}$ (figure~\ref{fig:correlationsH26}a), multiple-injection cases tend to yield somewhat larger $I_{\rm RE}$ than single injections. Since this comparison is made at fixed $N_{\mathrm{D,assim}}$, the difference cannot be attributed to a larger \emph{final} assimilated particle content. Consistently, identifying the corresponding points in figure~\ref{fig:correlationsH26}b indicates that, for similar $N_{\mathrm{D,assim}}$, $N_{\mathrm{Ne,assim}}$ is generally higher in the single-injection cases than in the multiple-injection cases. A plausible interpretation is that multiple injections may lead to a larger \emph{initial} assimilation early in the pre-TQ phase, accelerating the radiative collapse and shortening the pre-TQ, which limits subsequent ablation/assimilation and degrades the RE mitigation performance compared to single injections with comparable $N_{\mathrm{D,assim}}$.

Conversely, in the DTHmode24 scenario the plasmoid drift is significantly stronger, enhancing the outward drift of the ablated material. This tends to reduce the \emph{assimilation efficiency}. Nevertheless, because multiple injections supply a larger total inventory available for ablation, they still typically reach higher \emph{final} assimilated hydrogen content than single injections (figure~\ref{fig:correlationsDTH}a), which contributes to their improved mitigation performance, since the hydrogen recombination limit is not yet reached in these cases.

As noted already for single injection cases, the condition that the pre-TQ time is longer than the time it takes for the pellet to reach the plasma core must be satisfied for successful runaway suppression to be achieved. This is however not a sufficient condition, as a comparison between figures~\ref{fig:correlationsDTH}(c) and~\ref{fig:correlationsDTH-noact}(c) reveal. The main difference between these two figures is that in the former runaway generation via the Compton scattering and tritium decay mechanisms are accounted for, while they are neglected in the latter figure. When activated runaway sources are accounted for, the seed runaway current cannot be reduced sufficiently to suppress the runaway beam, even though the pre-TQ is long in these cases.

Staggered injections generally perform well, although they are not always the best performing cases. In the H26 L-mode scenario, the three staggered cases considered achieve by far the lowest runaway seed currents, which therefore lead to negligible runaway currents. This consistent success in runaway suppression is due to the long pre-TQ achieved by separating the injection of protium and neon in time, and enabled by the weak plasmoid drift in the relatively cold L-mode plasma. In contrast, in the DTHmode24 H-mode scenario, almost all of the protium in the first injection drifts out of the plasma, making the performance of the staggered scheme comparable to that of the single injection cases.

Successful mitigation across all simulations is strongly scenario dependent and can be summarized as follows. First, as shown in panel (a) of figures~\ref{fig:correlationsH26}, \ref{fig:correlationsDTH} and \ref{fig:correlationsDTH-noact}, the final runaway current scales inversely with the number of assimilated protium atoms. Thus, a high degree of assimilation of the pellet material must be achieved. Second, the pre-TQ must be extended for long enough that hot electrons can thermalize before the TQ is triggered. In the H26 L-mode scenario, this is best achieved by a staggered injection which reduces the plasma temperature with a first protium pellet, and then triggers the TQ with a second delayed neon pellet. In the DTHmode24 H-mode scenario, these conditions are best achieved by injecting multiple mixed protium-neon pellets. Due to the strong plasmoid drift of pure protium pellets, the mixture with neon ensures that the material is deposited within the plasma, and the multiple pellets raises the total number of atoms which are assimilated. Notably,  a key difference between the H26 L-mode and DTHmode24 H-mode scenarios is that, with our (late) TQ onset criterion, the H-mode cases typically require a longer time to reach the TQ. This extended pre-TQ duration allows pellet fragments to penetrate to inner radii and deposit material deeper in the plasma before the TQ.  We note however that it is only when nuclear runaway seed sources are disabled that runaway suppression can be achieved in DTHmode24.

\subsubsection{Effect of RE scrape-off}\label{sec:results:vde}
An important finding from the updated parameter scan is the role of runaway scrape-off losses caused by a VDE. We observe that in certain cases, especially those with near full runaway suppression, the inclusion of this loss model makes the difference between a small residual runaway current and the formation of a large runaway beam. For instance, in the L-mode H26 case S3 (single pellet injection, \(2.8\%\) Ne) discussed in section~\ref{sec:correlations}, the influence of the RE scrape-off model is dramatic. As already noted for this case, the low neon content of the pellet results in significant protium assimilation, a long pre-TQ, and consequently a small runaway seed before the TQ onset.
The small runaway seed population is then lost via radial diffusion during the relatively long transport event of the TQ. By the end of the TQ, only a negligible seed current 
survives. The subsequent avalanche multiplication is far too small to amplify this seed into a macroscopic current, as the plasma vertical motion and scrape-off of flux surfaces severely reduces the poloidal flux change within the confined volume.

\begin{figure}
    \centering
    \includegraphics[width=1\linewidth]{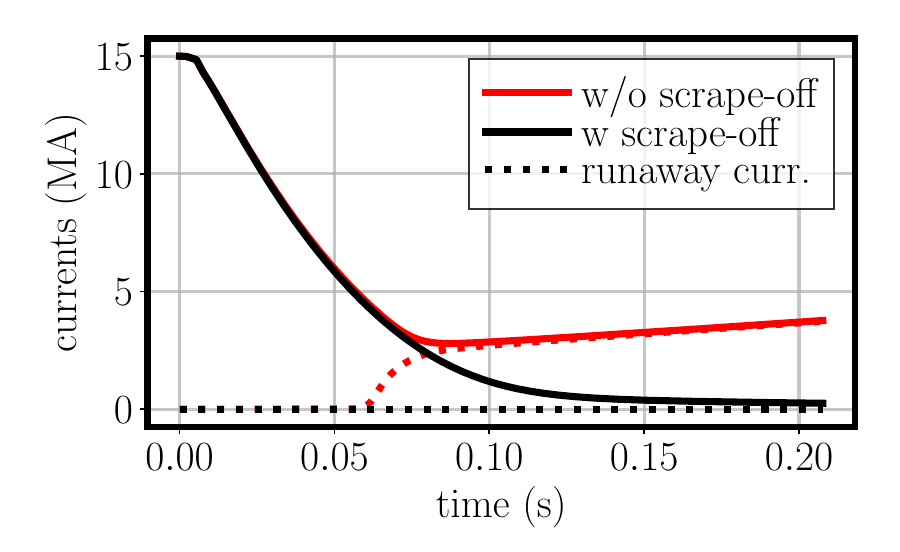}
    \caption{Evolution of the plasma current in the L-mode H26 case S3 with (black curves) and without (red curves) the RE scrape off losses. Dotted lines show the evolution of the runaway currents, while solid lines represent the total plasma current, calculated as the sum of the ohmic and the runaway current.}
    \label{fig:S3scrape-off}
\end{figure}

\begin{figure*}
    \begin{overpic}[width=1\linewidth]{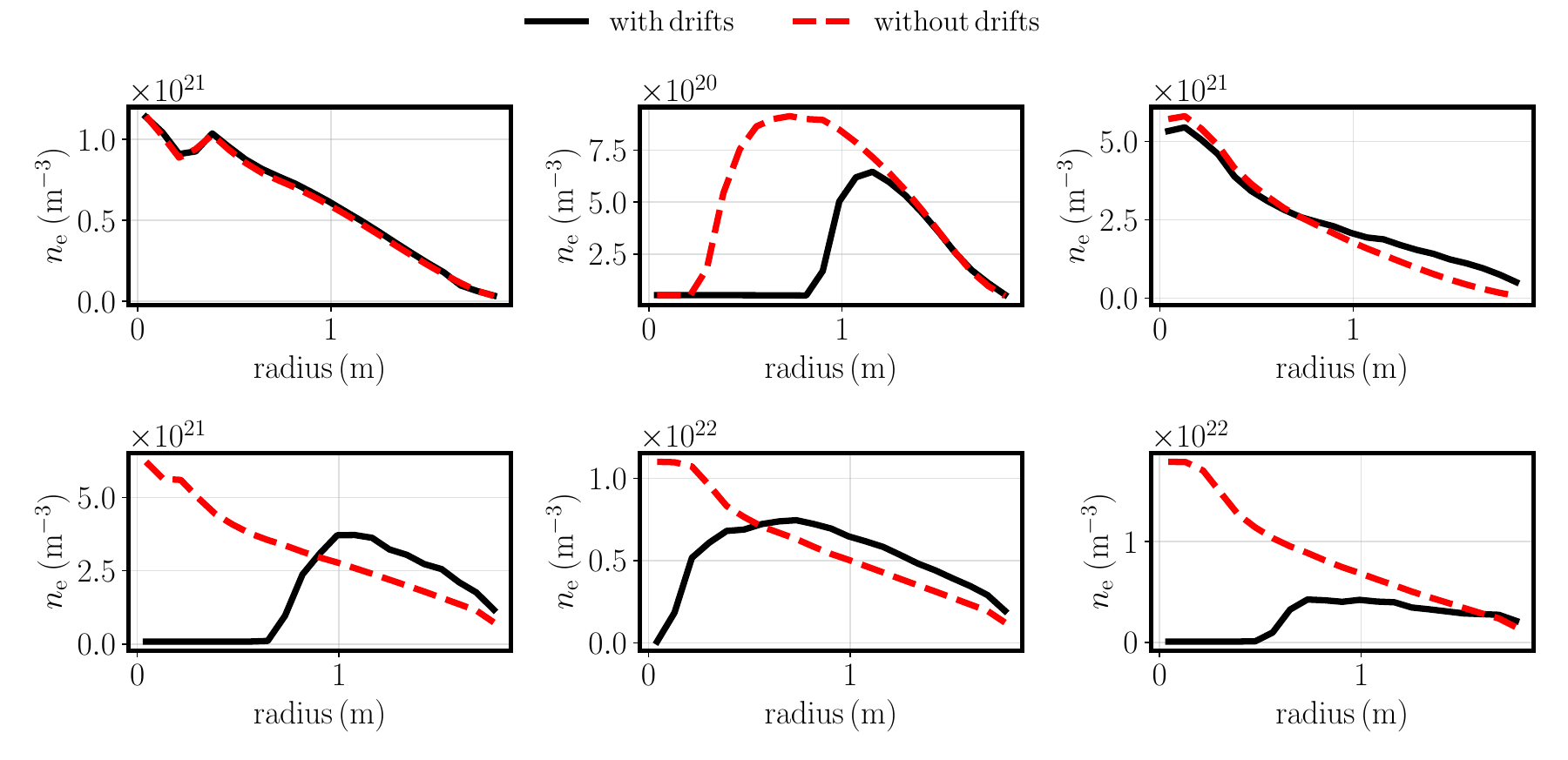}
    \put(26,39.6){\textbf{S2 (a)}}
    \put(57.6,39.6){\textbf{M2 (b)}}
    \put(90,39.6){\textbf{St3 (c)}}
    \put(24.6,17.6){\textbf{S12 (d)}}
    \put(57.6,17.6){\textbf{M6 (e)}}
    \put(90,17.6){\textbf{St6 (f)}}
    \end{overpic}
    \caption{Electron density profiles immediately before the thermal‑quench onset in six cases. Solid curves show simulations with plasmoid drift, while dashed curves those which do not include the drift model. In the low‑temperature L‑mode plasmas H26 (panels (a)-(c)) the drift broadens the profile only slightly except in case (b), but in the hotter H‑mode plasmas DTHmode24 (panels (d)-(f)) it drives a pronounced outward redistribution that lowers the core density by 10–30\%. Since the pre‑TQ densities set the collisional damping available to suppress the hot‑tail seed, this influences the final RE current. The staggered L‑mode case St3 retains a peaked, high‑density core and achieves near‑complete RE suppression, whereas its H‑mode counterpart St6 suffers a hollow profile which, combined with the presence of nuclear seeds, leads to a multi-megamapere RE beam.}
    \label{fig:drift_profiles}
\end{figure*}
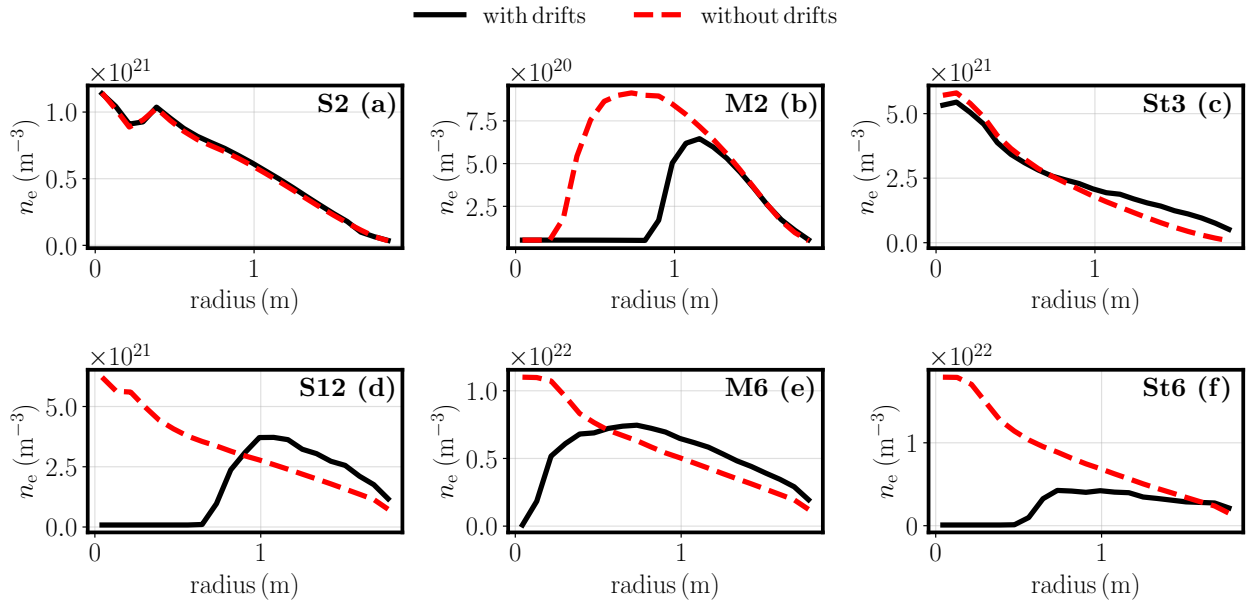

To isolate the effect of the scrape-off model, direct comparisons of simulations with and
without the scrape-off loss term are performed for cases near the RE suppression  as shown in figure~\ref{fig:S3scrape-off}. In the example above, both the scrape-off and
no-scrape-off simulations are initialized with nearly identical pre-TQ conditions and RE seed. However, in the no-scrape-off case the minuscule pre-TQ seed remains confined to the core and is accelerated by the post-TQ electric field, growing into a multi-MA runaway beam. In contrast, in the case with the scrape-off model the plasma gradually shrinks as the CQ evolves, therefore limiting the total avalanche gain that is achieved by reducing the available poloidal flux variation on each surface by the time the surface became open.

A similar behaviour to the S3 case is seen in several of the cases with complete runaway suppression. The scrape-off model limits the avalanche gain in cases with small runaway seeds. In cases with large runaway seeds, the scrape-off model is not as effective since a significant runaway current can then form before all flux surfaces are completely scraped off.

%%%%%%%%%%%%%%%%%%%%%%%%%%%%%%%%%%%%%%%%%%%%%%%%%%%%%%%%%%%%%%%%%%%%%%%%%%%%%%%%%%%%%%%%%%%%%%%%%%%%%%%%%%%%%%%%%%%%%%%%%%%%%%%%%%%%%%%%%%%%%%%%%%%%%%%%%%%%%%%%%%%%%%%%%%%%%%%%%%%%%%
\subsubsection{Effect of plasmoid drift}
\label{sec:plasmoiddrift}
The ablation cloud drift to the low-field side proves to be another critical factor that can undermine disruption mitigation. In general, we find that plasmoid drifts have a noticeable impact on the deposition profiles primarily for hydrogen-rich injections (especially pure D), whereas for neon-doped pellets the effect is relatively minor~\cite{Vallhagen2025b}. A pure D$_2$ pellet ablates into a hot ($T_e\sim 10$–\qty{50}{eV}) plasmoid with high internal pressure and minimal radiative cooling, which experiences a strong outward $E\times B$ force and can travel a significant distance towards the low field side before assimilating. In contrast, a neon-containing pellet produces a much cooler ablation cloud ($T_e\approx\qty{5}{eV}$ or less) because neon line radiation efficiently dissipates the plasmoid thermal energy, decreasing the pressure. The cooler, pressure-reduced plasmoid drifts more slowly and deposits its content closer to the original flux surface of ablation. In essence, the neon in the pellet “anchors” the material locally via radiative cooling, whereas hydrogenic fragments without impurities can rocket outward across flux surfaces.

In these simulations, the plasmoid temperature is prescribed to $30\,\mathrm{eV}$ for pure hydrogenic pellets and $5\,\mathrm{eV}$ for Ne-doped pellets. This parameter directly affects the total drift distance of the ablated material. We note that the characteristic plasmoid temperature is expected to be only weakly dependent on the background plasma temperature: to leading order, both the heat flux into the plasmoid and the ablation rate scale similarly with the upstream plasma conditions, so the effective energy balance that sets the plasmoid temperature varies little across the scenarios considered. It is therefore reasonable to use the same prescribed plasmoid temperatures in all scenarios, and the chosen values are consistent with typical measurements and modelling on smaller devices \cite{Vallhagen2025b}. 
The influence of plasmoid drift on material assimilation was quantified by performing paired simulations of each injection scheme, with the drift model alternately enabled (“drift on”) and disabled (“drift off”).
Figure~\ref{fig:drift_profiles} shows the resulting electron density \(n_e(r)\) at TQ onset for six representative cases: S2, M2, and St3 in the L-mode H26 scenario, and S12, M6, and St6 in the H-mode DTHmode24 scenario. In both scenarios the injected pellets contain at least a small neon fraction, except for the first stage in the staggered (St) cases, where the first injection consists only of protium.

\begin{figure*}
    \centering
    \begin{overpic}[width=0.9\linewidth]{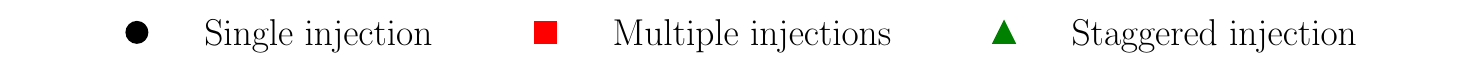}
    \end{overpic}

    \centering
    \begin{overpic}[width=1\linewidth]{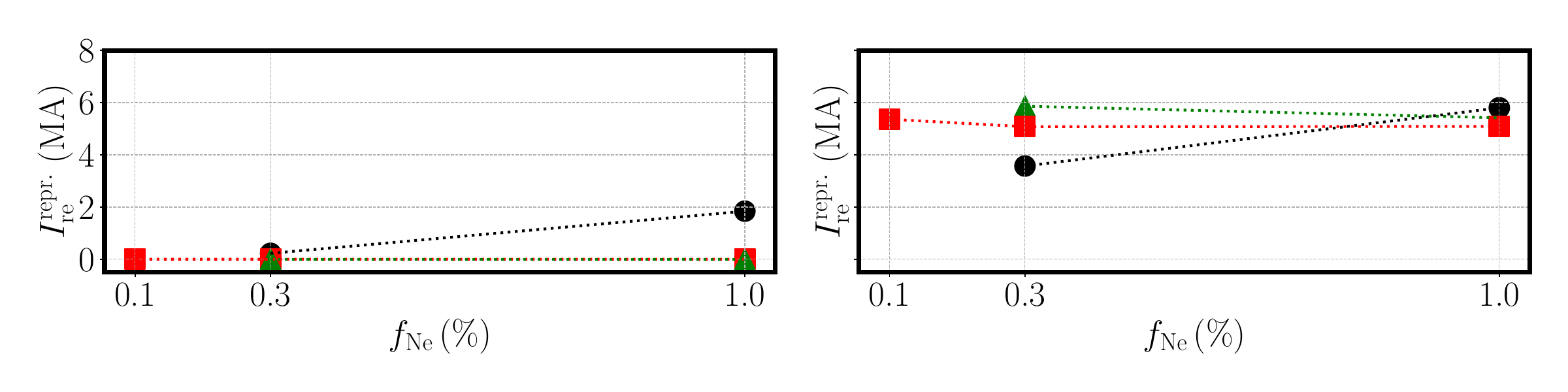}
    \put(23,23){\textbf{L-mode H}}
    \put(70,23){\textbf{H-mode DT}}
    \put(46,19.5){\textbf{(a)}}
    \put(94,19.5){\textbf{(b)}}
    \end{overpic}
    \caption{Representative runaway electron current $I_{\mathrm{re}}^{\mathrm{repr}}$ (in MA) as a function of the neon fraction $f_{\mathrm{Ne}}$ in the neon-containing pellet(s) (expressed as \% of total injected mass), for two ITER scenarios: (a) L-mode hydrogen plasmas and (b) H-mode deuterium-tritium plasmas. Different injection strategies are compared: single injection (black circles) S8 and S12, multiple injections (red squares) M2 and M6, and staggered injection (green triangles) St3 and St6. Decreasing the injected neon content consistenlty reduces the representative runaway current in all the non-nuclear cases.     }
    \label{fig:Nescan}
\end{figure*}

The most significant deviations between the drift- and no-drift cases is seen in the DTHmode24, figure~\ref{fig:Nescan}(d)-(f), where less material reaches the plasma core in the presence of drifts. In contrast, except for the case M2, the H26 scenario is only slightly affected by the plasmoid drifts. The stronger impact of drifts in DTHmode24 is not due to a different plasmoid temperature (which is assumed to be the same across scenarios), but rather to the fact that the hotter background plasma causes the pellets to ablate more peripherally. As a result, the material ablates closer to the edge, so that a given plasmoid drift displaces a larger fraction of the ablated material toward the low-field side and out of the confined region, reducing core penetration. Consequently, although the total material assimilation does not change significantly with and without the drifts in DTHmode24 (S12 and M6), the deposition location does. In both S12 and M6 more material is deposited near the edge because of the drifts, and very little if any material reaches the plasma core. Because of this, the temperature reduces faster in the outer parts of the plasma and triggers the TQ sooner when drifts are present. This in turn produces a larger hot-tail seed and representative RE current, since the faster edge cooling drives a more rapid global thermal quench while a substantial fraction of the core can still be hot and weakly collisional, i.e.\ before the injected material is fully homogenized by transport and increases the collisional drag in the core.

The DTHmode24 staggered injection case St6 similarly sees poor core penetration of material in the presence of drifts, but also has lower material assimilation overall. The reason for this is that the first pure D$_2$ pellet which is injected barely deposits any material inside the plasma, due to the significant plasmoid drifts, effectively rendering St6 a single injection case with a large Ne payload. Once the Ne pellet enters the plasma and starts depositing material, the TQ is quickly triggered and a significant hot-tail seed is generated.

Of the H26 cases in figure~\ref{fig:Nescan}(a)-(c), only M2 is significantly affected by the plasmoid drifts. In this case, the plasmoid drift causes material to accumulate in the outer parts of the plasma and triggers the TQ much earlier than in the case without drifts. The resulting material assimilation is much lower than without the drifts, and the short pre-TQ yields a large hot-tail seed. The other H26 cases see a negligible effect from the plasmoid drift. This reflects that M2 is a borderline case: a comparatively small drift-driven shift in the deposition towards the edge is sufficient to trigger the TQ substantially earlier, and once the TQ begins the remaining ablation (and hence further assimilation) is strongly reduced. Consequently, an order-unity change in the TQ onset time translates into a large change in the final assimilated inventory, and the shorter pre-TQ yields a larger hot-tail seed.

%%%%%%%%%%%%%%%%%%%%%%%%%%%%%%%%%%%%%%%%%%%%%%%%%%%%%%%%%%%%%%%%%%%%%%%%%%%%%%%%%%%%%%%%%%%%%%%%%%%%%%%%%%%%%%%%%%%%%%%%%%%%%%%%%%%%%%%%%%%%%%%%%%%%%%%%%%%%%%%%%%%%%%%%%%%%%%%%%%%%%%%%%%%%%%%%%%%%%%%%%%%%%%%%%%%%%%%%%%%%%%%%%%%%%%%%%%%%%%%%%%%%%%%%%%%%%%%%%%%%%%%%%%%%%%%%%%%%%%%%%%%%%%
\subsection{Low neon injection cases}
\label{sec:lowNe}

The experimental observation that pure protium SPI into strongly seeded plasmas can achieve longer pre-TQ durations and reduced radiation asymmetries, leading to enhanced RE avoidance, presents a novel pathway for thermal load mitigation during disruptions~\cite{Sheikh2025}. This suggests that investigating the efficacy of low neon injection, either as a component of the SPI pellet or through pre-existing plasma seeding, could be crucial for ITER, where the balance between effective energy radiation and the minimization of RE generation is essential.
To quantify the role of radiative losses during the CQ, some DTHmode24 and H26 reference simulations are repeated with progressively lower neon contents in the doped pellet, while keeping the total number of atoms per pellet fixed at $1.85 \times 10^{24}$. Specifically, simulations were carried out for neon fractions of $f_{\rm Ne} = 1.0\%$, $0.3\%$, and $0.1\%$.

Importantly, in our 1D simulations, where we do not account for the presence of intrinsic impurities in ITER plasmas, reduced neon radiative power tends to promote incomplete TQs. An incomplete TQ is here considered as a case in which a significant portion of the plasma is ohmically reheated to hundreds of eV after the initial TQ, leading to an unacceptably long CQ. To suppress the unphysical post-TQ reheating that can arise through unrealistically steep current-density gradients, the adaptive–hyperresistivity model discussed in section~\ref{sec:theory:hypres}, was activated in all simulations. The model successfully suppressed current spikes for every case with $f_{\rm Ne} \geq 0.3\%$ (apart from St3); at $f_{\rm Ne} = 0.1\%$ substantial reheating cannot be suppressed in S7, S8, S12, St3, and St6, because of the low Ne assimilation, and the consequent low radiation energy losses. The most demanding configuration was the early-onset S6 case (non-nuclear H26), where reheating could not be prevented at any reduced-neon level.

As shown in Fig.~\ref{fig:Nescan}, lowering the neon fraction in the injected pellet systematically reduces the representative runaway current in L-mode H26 plasmas. Physically, a reduced Ne content limits line radiation, thereby extending the pre–TQ and allowing the suprathermal electron tail to re-thermalize. The resulting hot-tail seed is suppressed as the hot-tail yield is exponentially sensitive to the TQ timescale. Additionally, reducing $f_{\mathrm{Ne}}$ increases the post disruption temperature, consequently reducing the induced parallel electric field. For a fixed total number of injected atoms, a lower Ne fraction also favors protium-driven densification, since the Ne tends to cool the plasma more and reduce the overall assimilation. This in turn tends to raise $E_{\rm c}$ and decrease $E_\parallel/E_{\rm c}$, further reducing avalanche growth.

Although this trend should in principle hold also for H-mode nuclear plasmas, we do not observe a uniform decrease of $I_{\rm RE}$ across all investigated cases. There are two main reasons: first, reducing the neon content lowers radiative cooling during the ablation/deposition phase, so the background plasma remains hotter and material tends to start ablating more peripherally. This enhances an edge localized drift which in turn reduces core assimilation and limits the pre-TQ density rise; consequently, the benefit of a longer pre-TQ is partly offset. Note that in our semi-analytical plasmoid-drift model, the drift drive is not made composition-dependent, so varying $f_{\mathrm{Ne}}$ does not directly strengthen the drift.
 Second, and more importantly, even when the hot-tail seed is mitigated, DT operation is dominated by an intrinsic Compton scattering runaway source driven by intense $\gamma$-ray fluxes during the disruption. This Compton seed is only weakly affected by variations in the Ne fraction and, for ITER-relevant DT conditions, it is sufficiently strong to produce a multi-megaampere runaway beam through avalanche multiplication. Consequently, in DT H-mode scenarios the advantage of low $f_{\mathrm{Ne}}$ is limited by the large Compton-generated seed, so that reducing the neon fraction alone is insufficient to suppress the runaway current.

%%%%%%%%%%%%%%%%%%%%%%%%%%%%%%%%%%%%%%%%%%%%%%%%%%%%%%%%%%%%%%%%%%%%%%%%%%%%%%%%%%%%%%%%%%%%%%%%%%%%%%%%%%%%%%%%%%%%%%%%%%%%%%%%%%%%%%%%%%%%%%%%%%%%%%%%%%%%%%%%%%%%%%%%%%%%%%%%%%%%%%%%%%%%%%%%%%%%%%%%%%%%%%%%%%%%%%%%%%%%%%%%%%%%%%%%%%%%%%%%%%%%%%%%%%%%%%%%%%%%%%%%%%%%%%%%%%%%%%%%%%%%%%
\subsection{Effect of hyper-resistivity during TQ}
\label{sec:hypresTQ}

\label{sec:TQhypres}

Previous studies \cite{Boozer2018, Nardon2023} have shown that the TQ induces a rapid relaxation of the plasma current profile, often accompanied by a transient spike in total plasma current. Boozer \cite{Boozer2018} has pointed out that the abrupt drop in internal inductance and the observed TQ current spike imply a near-complete destruction of magnetic flux surfaces, allowing the current-density profile to flatten on sub-millisecond timescales. This fast reconnection process far outpaces classical resistive diffusion, underlining the importance of turbulent magnetic stochasticity in the TQ. Recent JOREK simulations by Nardon \emph{et al.} \cite{Nardon2023} confirm this picture: as the 2/1 tearing mode grows and field lines become stochastic, the toroidal current $j_\phi$ redistributes into a broad, flattened profile within $\sim\qty{1}{ms}$.

To model these effects in a 1D disruption simulation framework, such as \DREAM, we include a hyper-resisitive term in the Faraday's law \cite{Pusztai2022}
\begin{equation}
\frac{\partial\psi_{\rm p}}{\partial t}
=-\mathcal{R}+
\underbrace{\mu_0\frac{\partial}{\partial\psi_{\rm t}}\left(
      \psi_{\rm t}\Lambda_{\rm m}\frac{\partial}{\partial\psi_{\rm t}}
      \frac{j_{\mathrm{tot}}}{B}\right)}_{\text{hyper-diffusion}},
      \label{eq:hypres}
\end{equation}
where $\mathcal{R}\equiv 2\pi\langle \mathbf{E}\!\cdot\!\mathbf{B}\rangle/\langle \mathbf{B}\!\cdot\!\nabla\varphi\rangle$ is the resistive (inductive) drive, equal to the local loop voltage when the second term vanishes, and $\varphi$ is the toroidal angle. The toroidal magnetic flux $\psi_{\rm t}$ is used as radial coordinate and $\Lambda_{\rm m}$ is the magnetic helicity transport coefficient, sometimes referred to as the hyper-resisitivity. Simulations without this term are unable to produce the observed current density profile flattening or current spike, when using Spitzer resistivity alone. The added hyper-resistivity $\Lambda_{\rm m}$ serves as a magnetic helicity transport coefficient that mimics the MHD reconnection taking place in the real plasma. Specifically, as shown in Eq.~\ref{eq:hypres}, the hyper-resistive contribution appears as a radial hyper-diffusion operator acting on $j_{\rm tot}/B$. This term effectively redistributes toroidal current density across flux surfaces on a rapid timescale. 

Boozer \cite{Boozer2018} also derived a heuristic connection between $\Lambda_{\rm m}$ and the underlying field-line chaos, given in equation~\eqref{eq:LambdaHypres}, finding that $\Lambda_{\rm m}\propto \Delta r_{\rm s}^2/N_{\rm t}$ with $\Delta r_{\rm s}$ denoting the radial width of the stochastic region, and $N_{\rm t}$ the number of toroidal transits a field line needs to wander across that region. As discussed in Sec.~\ref{sec:physicsmodels}, this relationship is analogous to the Rechester–Rosenbluth model for turbulent heat conduction and implies that a higher level of magnetic perturbation corresponds to a larger effective hyper-resistivity. In practice, this guides the choice of $\Lambda_{\rm m}$ in our 1D simulations. By calibrating $\Lambda_{\rm m}$ to a value consistent with fast current-profile relaxation (e.g.\ $\Lambda_\mathrm{m}\approx\qty{6.3e-1}{\weber\squared\per\meter\second}$, as in~\cite{Vallhagen2025}), assuming a current spike equal to $\sim 2.5 \%$ of the initial plasma current $I_{\rm p}$, across the simulated cases. 

A subtlety arises when combining the hyper-resistive model with runaway-electron kinetics. In the mean-field formulation of~\cite{Boozer1986,Boozer2018}, the total parallel electric field contains both a resistive and a hyper-resistive contribution, \(E_{\parallel} = E_{\mathrm{res}} + E_{\mathrm{hyp}}\). A direct evaluation of runaway generation rates at the total field \(E_{\parallel}\) is however problematic: the hyper-resistive term can drive strong local electric fields, including sign reversals, that cause numerical instabilities and, more fundamentally, the hyper-resistive contribution to the electric field is not included in the standard kinetic equation used to evaluate runaway rates. As discussed by Boozer~\cite{Boozer2019}, the hyper-resistive term should enter the kinetic equation as a separate operator reflecting the stochastic field-line transport, and its divergence structure ensures that it redistributes, but does not create, magnetic helicity. Consequently, an electron stochastically sampling the reconnection region experiences no net acceleration from the hyper-resistive contribution~\cite{Boozer2019}. This supports the approach adopted here, in which the runaway generation rates are evaluated using only the resistive component of the electric field, \(E_{\mathrm{res}}\), which is the part responsible for net energy transfer to electrons. A fully self-consistent kinetic treatment incorporating the hyper-resistive operator à la Boozer~\cite{Boozer2019} remains an avenue for future work; however, the present
approach is expected to capture the dominant effect on runaway dynamics, namely the
reduction of the resistive poloidal-flux change discussed later.

\begin{figure}    \centering
    \begin{overpic}[width=1\linewidth]{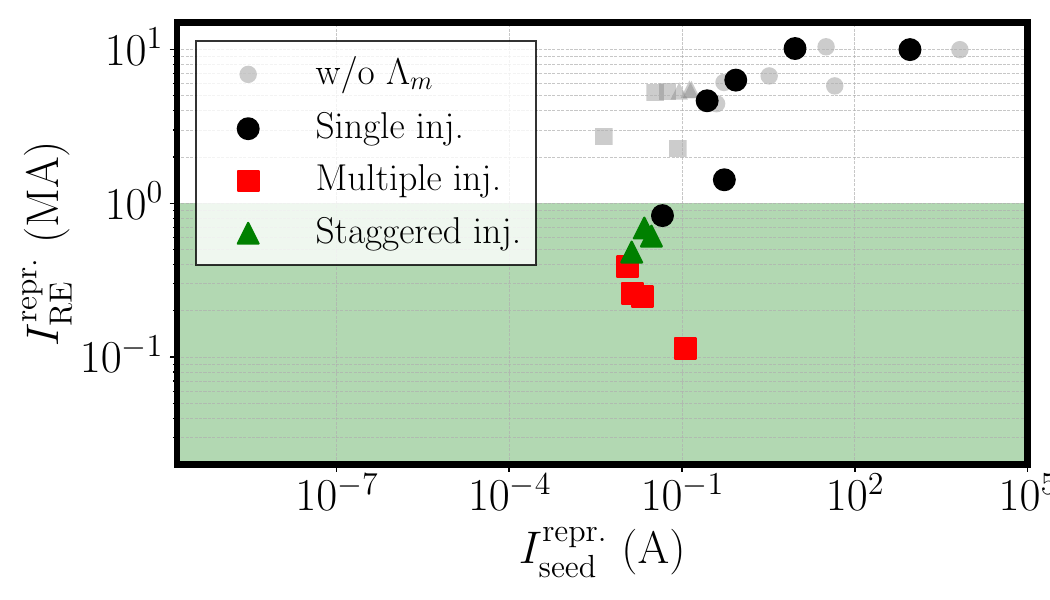}
                     \put(80,38){\color{black}{\textbf{1 MA}}}
    \end{overpic}
    \caption{Correlation between the representative runaway electron current $I_\mathrm{RE}^\mathrm{repr.}$ and the representative seed current $I_\mathrm{seed}^\mathrm{repr.}$ for ITER DT H-mode (DTHmode24) scenarios, comparing cases \emph{with} and \emph{without} hyper-resistivity during the transport event. Different symbols denote the injection scheme: single-pellet injection (circles), multiple simultaneous pellet injections (squares) and staggered injections (triangles). Gray symbols
    correspond to simulations without the hyper-resistive term ($\Lambda_{\mathrm{m}}=0$), while the colored symbols show results with a uniform hyper-resistivity ($\Lambda_{\mathrm{m}}=\qty{0.628}{\weber\squared\per\meter\second}$) applied during the TQ. With hyper-resistive transport included, the final runaway-electron current is reduced by roughly an order of magnitude across all the high-assimilation cases, despite the similar seed magnitudes. The region below $I_{\rm RE}^{\rm repr.}=\qty{1}{MA}$, indicated in green, shows the range of tolerable runaway current, when the when impact conditions are favorable.}
    \label{fig:HypresAll}
\end{figure}

To assess the impact of hyper-resistive transport, we repeated the DT H-mode baseline cases, this time applying a spatially uniform hyper-resistivity  \(\Lambda_{\mathrm{m}}\) during the transport event. In such cases, although the seed accumulated by the end of the TQ is essentially unchanged relative to the \(\Lambda_{\mathrm{m}}=0\) simulations, the representative runaway current decreases by one order of magnitude across all cases where the representative RE seed current remains between $10^{-2}$ to $10^{-1}\,\si{A}$ as shown in Fig.~\ref{fig:HypresAll}.
This reduction brings the post-CQ runaway current below $\sim$1~MA in these cases, which may be more manageable, although tolerability ultimately depends on the impact characteristics and is not assessed here.

\begin{figure}
    \centering
    \includegraphics[width=0.85\linewidth]{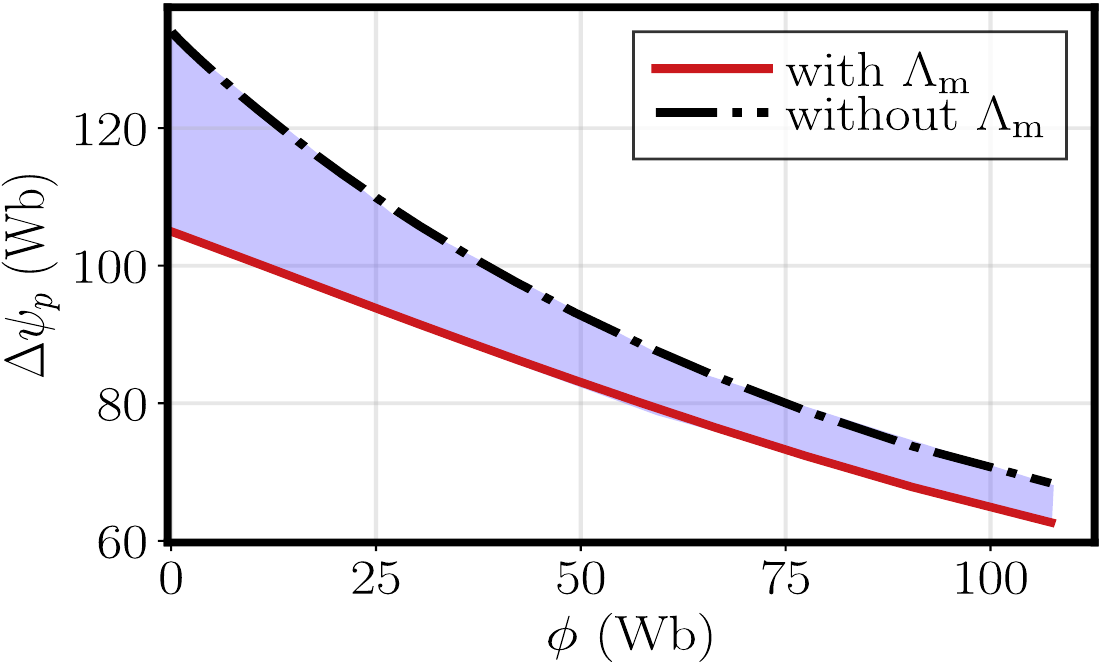}
    \caption{Absolute variation of the poloidal magnetic flux, $\Delta\psi_{\rm p}$, during the CQ versus the toroidal magnetic flux $\phi$ in the DT H-mode case M6 ($\phi=0$ at the magnetic axis, increasing towards the edge). Including hyper-resistivity during the transport event (red solid) reduces $\Delta\psi_{\rm p}$ across the radius relative to the reference without $\Lambda_{\rm m}$ (black dash-dotted), and most strongly in the core. This reduced poloidal-flux drop limits the avalanche gain which is the primary cause of the negligible representative runaway current when $\Lambda_{\rm m}$ is included.}
    \label{fig:hypMech}
\end{figure}

The primary mechanism for runaway current suppression via hyper-resistivity is the observed reduction in the poloidal-flux variation, $\Delta\psi_p$. Although the magnitude of this reduction is relatively modest  around 20\% in the core for the M6 scenario, as shown in Fig.~\ref{fig:hypMech}, its consequences for avalanche multiplication can be substantial since the avalanche gain depends exponentially on $\Delta\psi_{\rm p}$~\cite{Boozer2018}. 

We emphasize that the ratios inferred from figure~\ref{fig:HypresAll} correspond to a representative gain for the finite seed currents present in those simulations, and therefore should not be interpreted as the potential avalanche multiplication (i.e.\ the limit as the seed tends to zero). To estimate the impact on the potential gain, we use the scaling reported by Wang \emph{et al.}~\cite{Wang2024}, where an increase of $\Delta\psi_p \simeq 5\,\mathrm{Wb}$ corresponds approximately to one ten fold increase in avalanche gain. Since hyper-resistivity reduces the central $\Delta\psi_p$ by about $\qty{20}{Wb}$ in figure~\ref{fig:hypMech}, the potential avalanche gain is expected to decrease by
\begin{equation}
  \frac{G_{\mathrm{ava}}^{(\Lambda_m)}}{G_{\mathrm{ava}}^{(0)}} \sim 10^{-\,\Delta\psi_p/(5\,\mathrm{Wb})}
  \approx 10^{-20/5} \approx 10^{-4}.
\end{equation}
For the finite-seed cases in figure~\ref{fig:HypresAll}, the representative gain decreases more mildly, but the reduction in potential gain is the more relevant metric when assessing the propensity for avalanche multiplication from very small seeds.

%%%%%%%%%%%%%%%%%%%%%%%%%%%%%%%%%%%%%%%%%%%%%%%%%%%%%%%%%%%%%%%%%%%%%%%%%%%%%%%%%%%%%%%%%%%%%%%%%%%%%%%%%%%%%%%%%%%%%%%%%%%%%%%%%%%%%%%%%%%%%%%%%%%%%%%%%%%%%%%%%%%%%%%%%%%%%%%%%%%%%%%%%%%%%%%%%%%%%%%%%%%%%%%%%%%%%%%%%%%%%%%%%%%%%%%%%%%%%%%%%%%%%%%%%%%%%%%%%%%%%%%%%%%%%%%%%%%%%%%%%%%%%%
\subsection{Effect of runaway transport during CQ}
\label{sec:CQdBB}
In all the disruption simulations discussed above, the runaway electron population is assumed to remain perfectly confined during the CQ. This assumption could underestimate the runaway electron losses during this phase. Strong MHD activity in the CQ can lead to stochastic magnetic fields, causing radial transport and loss of runaways. As shown in previous studies~\cite{MartinSolis2021}, sufficiently large magnetic perturbations during the CQ can completely suppress avalanche multiplication in ITER-like scenarios, even if the direct runaway scrape-off effect is neglected. Here, we extend our model to examine how introducing a Rechester–Rosenbluth-type diffusive transport of REs during the CQ affects the avalanche gain and final runaway current.

The runaway loss rate due to transport in the CQ is calculated as
\begin{equation}
    \left(\frac{\partial n_\mathrm{re}}{\partial r}\right)_\mathrm{transp.} = - \frac{1}{V'}
\frac{\partial}{\partial r}\!\left(
  V'\,D_{\rm re}\,\frac{\partial n_{\mathrm{re}}}{\partial r}
\right),
\end{equation}
where $V'$ is the flux-surface volume element and the RE transport coefficient $D_{\rm re}$ is calculated as in \eqref{eq:rechesterRosenbluth}, employing a fixed safety factor of $q = 1$. This choice provides a conservative estimate of transport losses, since in reality $q$ varies with radius and time during the CQ. From the diffusion coefficient, the radial loss time for a runaway electron can be estimated as $\tau_{\text{loss}} = a^2/(2D_{\rm re})$, as the characteristic time for a RE to diffuse from the core (radius $r=0$) to the edge ($r=a$).

\begin{figure}
    \centering
    \includegraphics[width=1\linewidth]{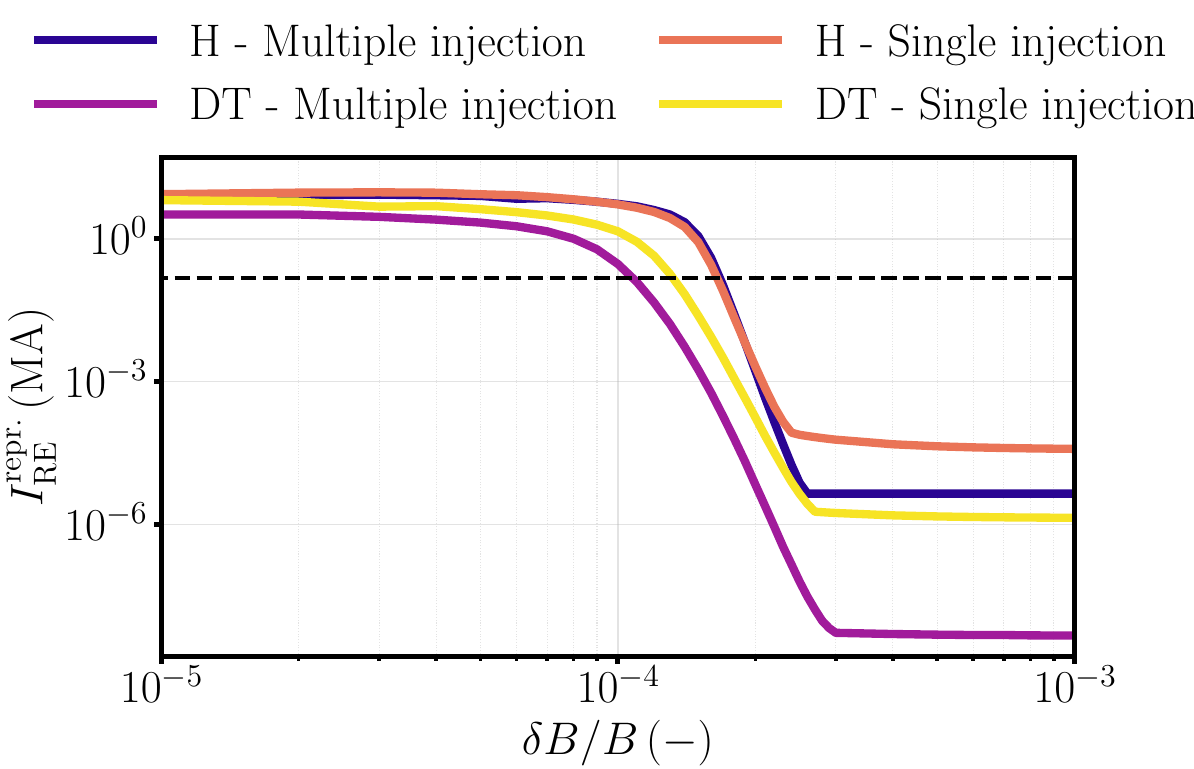}
    \caption{Representative runaway electron current for different magnetic perturbation levels. The investigated cases are some of the most challenging for runaway mitigation: M4 in blue (L-mode H multiple injections), M8 in purple (DT H-mode multiple injections), S6 in orange (L-mode H single injection), S15 in yellow (DT H-mode single injection). The dashed line denotes the \qty{150}{kA} current level.}
    \label{fig:dBBAll}
\end{figure}

To test the sensitivity of the final runaway current to the CQ transport, we varied the magnetic perturbation level $\delta B/B$ between $10^{-5}$ and $10^{-3}$. The corresponding diffusion coefficient thus varies from $\qty{0.6}{m^2/s}$ to approximately $\qty{6000}{m^2/s}$. In reality, the CQ perturbation is unlikely to be spatially uniform: $\delta B / B$ can vary in radius and time, and residual magnetic islands may allow localized confinement of REs. Here, for simplicity, we assume a spatially uniform $\delta B / B$ (and corresponding diffusion coefficient) and do not model any island-induced confinement regions or transport barriers. Figure \ref{fig:dBBAll} summarizes the impact of varying magnetic perturbation levels on the final RE current in several of our most challenging scenarios in terms of RE mitigation. We selected four cases that, with no perturbations, yield large runaway currents: a multiple-injection case in the L-mode H plasma (M4), a multiple-injection case in the H-mode DT plasma (M8), a single-pellet injection case in L-mode H (S6), a single-pellet case in H-mode DT (S15). The curves in Fig.~\ref{fig:dBBAll} show a clear threshold behavior in $I_{\text{RE}}$ as $\delta B/B$ increases. For weak perturbations ($\delta B/B \lesssim 10^{-4}$, corresponding to $D_{\text{re}}\lesssim 60$ m$^2$/s), the diffusion is too small to significantly alter the outcome, all these scenarios still produce multi-MA runaway currents. In this regime, the characteristic loss time $\tau_{\text{loss}}$ is comparable with the duration of the CQ, meaning REs are not expelled fast enough to surpass the avalanche multiplication. However, once the perturbation strength exceeds a critical value on the order of a few $10^{-4}$, the final runaway current in each case begins to drop sharply. By $\delta B/B \sim 4\times10^{-4}$ ($D_{\text{re}}\sim\qty{94}{m^2/s}$), the RE current is suppressed in all the cases studied, falling well below the \qty{150}{kA} limit indicated by the dotted line in Fig.~\ref{fig:dBBAll}. In fact, beyond this threshold, increasing $\delta B/B$ further had little additional effect as the runaway current was already reduced to a minimal level in every scenario.

\begin{figure}
    \centering
    \includegraphics[width=0.91\linewidth]{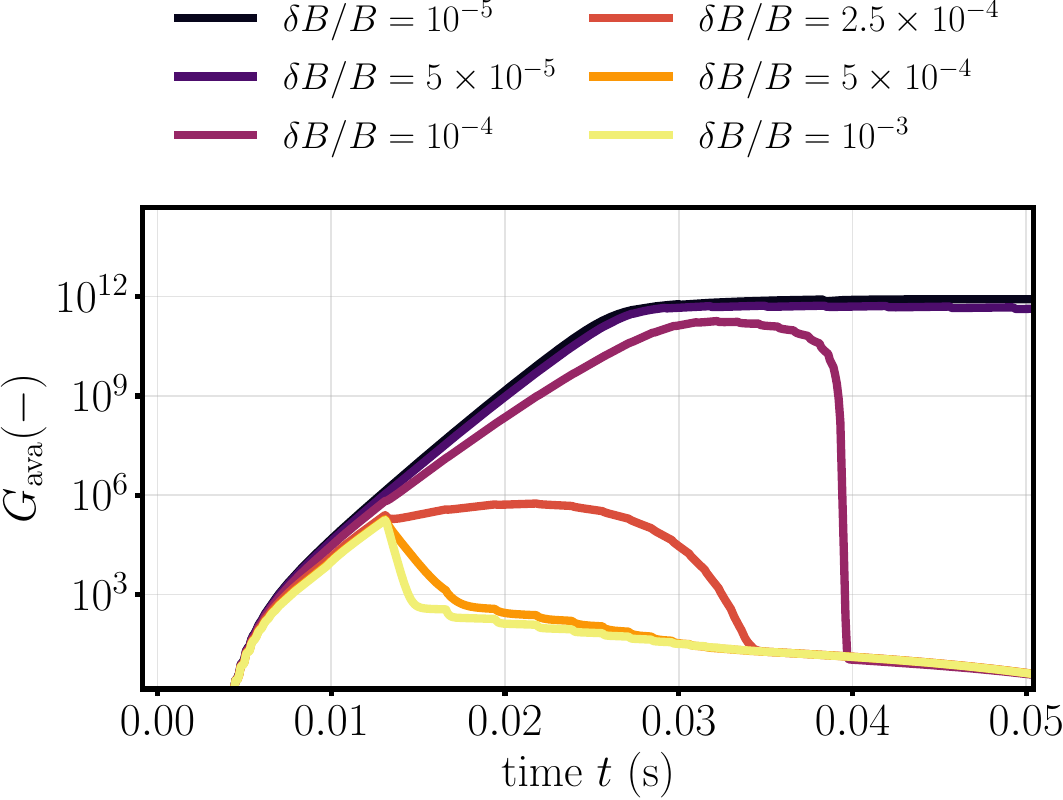}
    \caption{Time trace of the avalanche gain $G_{\mathrm{ava}}(t)=I_{\mathrm{RE}}(t)/I_{\mathrm{seed}}$ during the CQ phase of the case S15. Six values of the  magnetic perturbation amplitude are considered, spanning $\delta B/B = 10^{-5}$–$10^{-3}$. A weak perturbation ($\delta B/B\lesssim10^{-4}$) allows unconstrained avalanche growth to $G_{\mathrm{ava}}\sim10^{11}$ within $\sim\qty{40}{ms}$, whereas stronger perturbations progressively reduce the gain and precipitates an earlier collapse of the runaway current.}
    \label{fig:S15dBB}
\end{figure}

To illustrate the dynamics behind this threshold, Fig.~\ref{fig:S15dBB} shows time traces of the avalanche multiplication factor $G_{\text{ava}}(t) = I_{\text{RE}}(t)/I_{\text{seed}}$ during the CQ of the representative case S15, under different perturbation amplitudes. 

With very weak perturbations ($\delta B/B \sim10^{-5}$), the avalanche proceeds essentially unaffected, in fact the gain reaches $G_{\text{ava}}\sim10^{11}$ within \qty{40}{ms}. As $\delta B/B$ is increased to $1$–$3\times10^{-4}$, the avalanche growth is dramatically damped. The peak multiplication factor is much lower, and the RE current saturates and decays at an earlier time. This corresponds to the regime where $\tau_{\text{loss}}$ becomes shorter than the avalanche growth time, i.e.\ radial transport is fast enough to deplete runaways as they are trying to multiply. In practical terms, this suggests that if one can induce a strong magnetic perturbation during the CQ (for instance external 3D fields that induce a short-wavelength perturbation of order $\delta B/B \gtrsim 4\times10^{-4}$), it could serve as a complementary strategy to reduce the avalanche gain to tolerable levels. Such perturbations, induced by for example resonant magnetic perturbation (RMP) coils, have previously been studied extensively as a primary runaway avoidance technique, with limited success due to the inability of the perturbations to stochastize the magnetic field deep in the plasma~\cite{Yoshino2000,Lehnen2008,Papp2011}. We note that when combined with the VDE scrape-off effect discussed in sections~\ref{sec:theory:vde} and~\ref{sec:results:vde} RMPs may prove to be effective in suppressing runaway electrons. In fact, when a VDE is present, enhanced RE losses caused by externally applied 3D fields can in principle couple back to the VDE dynamics: additional stochastic losses reduce the RE current, and thus the poloidal flux carried by the RE beam, which can modify the vertical force balance and accelerate the vertical motion of the remaining current channel, further increasing RE losses. Moreover, as the vertical motion progresses, the distance between the plasma edge and the runaway current channel decreases, reducing the effective penetration depth required for the perturbations to affect the beam region and making it more plausible for the magnetic field to become stochastic there (although plasma-response screening may still limit the penetration).

\begin{figure*}
    \centering
        \begin{overpic}[width=.7\linewidth]{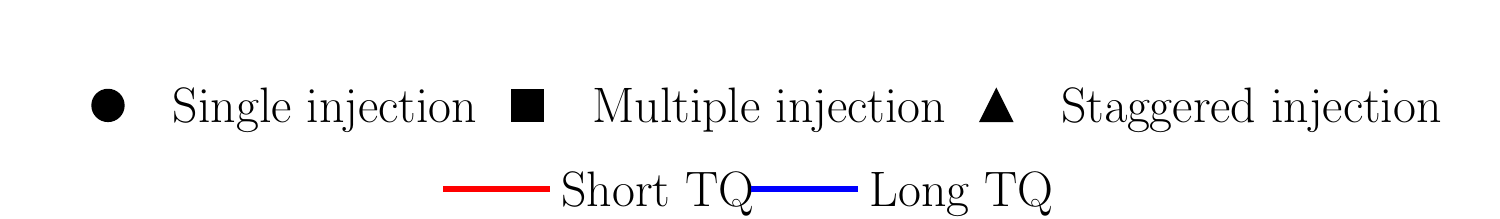}       
    \end{overpic}
    \vspace{+1mm}
    
        \begin{overpic}[width=1\linewidth]{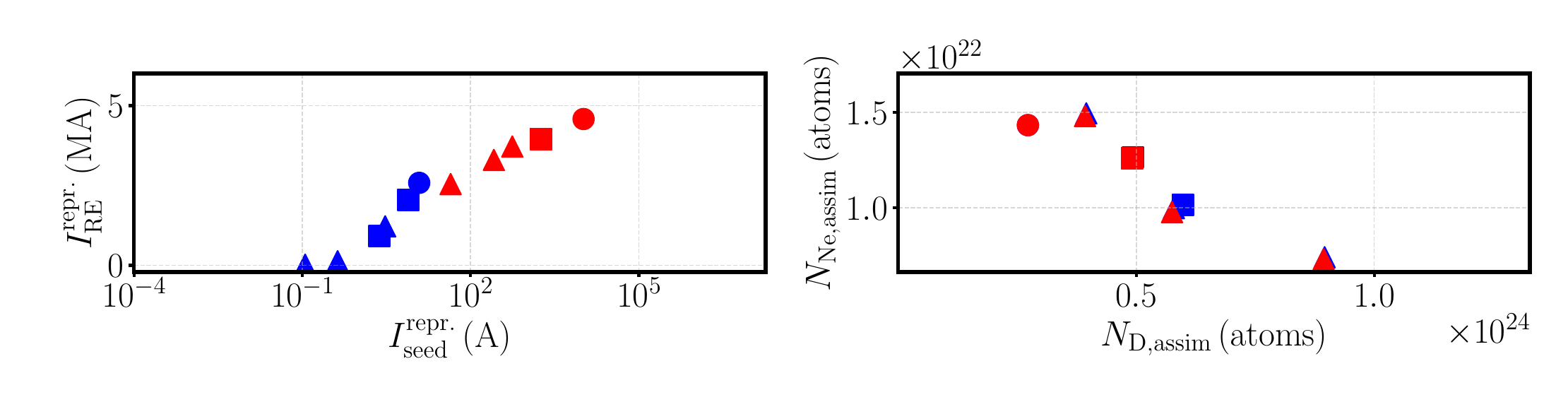}
    \put(9,19){\textbf{(a)}}
    \put(58,19){\textbf{(b)}}
    \end{overpic}%
                
    \caption{Runaway electron mitigation performance in the reduced-current ($I_{\rm p} = \qty{7.5}{MA}$) DD7.5MA scenario. Panel~(a) shows the representative runaway electron current $I_{\mathrm{RE}}^{\mathrm{repr}}$ as a function of the representative seed current $I_{\mathrm{seed}}^{\mathrm{repr}}$. Panel~(b) shows the number of assimilated neon atoms $N_{\rm Ne,\mathrm{assim}}$ as a function of the number of  assimilated protium atoms $N_{\rm H,\mathrm{assim}}$. Marker shapes denote the injection strategy: single-pellet (circles), multiple-pellet (squares), and staggered (triangles). Marker colors distinguish between long (blue, $t_{\mathrm{TQ}} = \qty{3}{ms}$) and short (red, $t_{\mathrm{TQ}} = \qty{1}{ms}$) TQ conditions. The results demonstrate that only the combination of a staggered injection strategy (triangles) and favourable TQ conditions (blue), which maximizes protium assimilation and minimizes the surviving seed current, leads to complete runaway electron suppression.}
    \label{fig:ddire_analysis}
\end{figure*}

%%%%%%%%%%%%%%%%%%%%%%%%%%%%%%%%%%%%%%%%%%%%%%%%%%%%%%%%%%%%%%%%%%%%%%%%%%%%%%%%%%%%%%%%%%%%%%%%%%%%%%%%%%%%%%%%%%%%%%%%%%%%%%%%%%%%%%%%%%%%%%%%%%%%%%%%%%%%%%%%%%%%%%%%%%%%%%%%%%%%%%%%%%%%%%%%%%%%%%%%%%%%%%%%%%%%%%%%%%%%%%%%%%%%%%%%%%%%%%%%%%%%%%%%%%%%%%%%%%%%%%%%%%%%%%%%%%%%%%%%%%%%%%
\subsection{Scenario with reduced plasma current}
\label{sec:reducedCurrent}

We now turn our attention to a non-nuclear case with a reduced plasma current of $I_\mathrm{p} = \qty{7.5}{MA}$, referred to in this work as \emph{DD7.5MA}. The primary motivation for studying a lower-$I_\mathrm{p}$ disruption is to assess how the RE generation dynamics and mitigation requirements scale with plasma current. Intuitively, a lower pre-disruption current means that less total poloidal flux is available to drive the avalanche multiplication, potentially relaxing the requirements for runaway avoidance. On the other hand, some mitigation strategies may be less effective at lower current if, for example, the plasma parameters or shard penetration depths differ. Here we simulate disruption mitigation in the \qty{7.5}{MA} ITER plasma using the same set of models (including drifts, scrape-off, etc.), and compare the results with the findings at \qty{15}{MA}. Notably, the nuclear seed sources are neglected in what follows.

We investigated three pellet injection strategies for the 7.5 MA disruption mitigation. The first strategy uses a single protium pellet doped with a large neon content ($N_{Ne} = 9.5\times10^{22}$ atoms) to trigger a rapid radiative collapse. The second strategy injects multiple pellets simultaneously (here 2 or 3 pellets), keeping the total neon inventory equal to that of the single injection.
The third strategy is a staggered injection. An initial injection of one to three nearly pure-D pellets with only a minimal neon doping ($N_{Ne}\approx1.85\times10^{21}$ atoms each, $\sim0.1\%$ neon fraction) to prolong the pre-TQ phase, followed by a second injection of a neon-doped H pellet identical to the single-pellet case (with $9.5\times10^{22}$ Ne atoms) . For this analysis, we assume a late TQ onset criterion and we evaluate each injection scheme under two distinct TQ duration conditions: a favorable case with a longer TQ of $t_{\rm TQ}=\qty{3}{ms}$, and an unfavorable case with a shorter TQ of $t_{\rm TQ}=\qty{1}{ms}$. It is worth stressing that the present simulations do not include hyper-resistivity during the TQ phase, nor any explicit modeling of runaway transport during the CQ. 

The results, presented in figure~\ref{fig:ddire_analysis}, reveal a clear hierarchy in the effectiveness of the tested strategies. Single pellet injection and simultaneous multiple injections were unable to suppress  runaway formation, irrespective of the TQ conditions. The staggered injection strategy, by contrast, was able to fully suppress runaway generation, but only in the favorable TQ scenario (\qty{3}{ms} TQ duration). When the same staggered scheme was applied with an unfavorable, rapid TQ of \qty{1}{ms}, it also failed to suppress the formation of a multi-MA RE beam. In summary, complete RE suppression in the \qty{7.5}{MA} plasma was achieved only within a narrow operational window: using a staggered SPI and having a sufficiently long TQ.

The outcomes of the simulations are summarized in Fig.~\ref{fig:ddire_analysis}. Panel (a) shows the representative RE current vs.\ the representative seed current for all the cases, while panel (b) shows the number of assimilated neon atoms vs.\ protium atoms. The staggered-injection cases (triangle markers) combined with a long \qty{3}{ms} TQ (blue points) cluster near zero runaway current, whereas all other cases end with a high RE current. This is consistent with the finding for the \qty{15}{MA} baseline cases that multiple favourable conditions must coincide to avoid runaways.

The success of the staggered injection cases in long TQ disruptions stems from two physical features. The first is the longer pre-TQ enabled by the staggered scheme, which first dilutes the plasma with a large amount of D and allows most of the hot electrons to thermalize. We note that since DD7.5MA is an H-mode scenario with a high temperature, it suffers from the ``H-mode penalty'' described in section~\ref{sec:plasmoiddrift} which causes the contents of pure protium pellets to drift towards the low-field side. To prevent this drift, the staggered injections considered here were doped with 0.1\% Ne, which is assumed to be sufficient to cool the ablation cloud and reduce the plasmoid drift, without prematurely triggering a radiative collapse. The second physical feature contributing to the success of the RE mitigation is the long TQ duration which gives sufficient time for the enhanced RE transport to eliminate any surviving hot-tail seed electrons from the pre-TQ.

A key conclusion emerges when comparing these \qty{7.5}{MA} results with the \qty{15}{MA} scenarios. The RE current conversion efficiency is markedly lower at \qty{7.5}{MA}, yielding a more favourable mitigation environment. This is evident in the tolerance to the seed current: as shown in Figure~\ref{fig:ddire_analysis}, a representative seed current of up to $\sim \qty{0.4}{A}$ can be robustly suppressed at \qty{7.5}{MA}, in sharp contrast to the \qty{15}{MA} scenarios, where successful mitigation required the representative seed to be effectively zero.

Finally, in every case of successful runaway current suppression, it is important to note that scrape-off losses of runaway electrons play a decisive role. In fact, the staggered injection cases that are able to fully suppress runaways achieve this outcome only because the scrape-off rate exceeds the avalanche growth rate. This condition is facilitated by the combination of a reduced representative seed population and a high post-injection electron density, with the low seed playing the dominant role. 
Notably, in this scenario we adopt a comparatively large neon content in the H/Ne pellets. This choice is partly pragmatic: within the plasmoid-drift model, the reduced magnetic field of this scenario leads to a substantially larger plasmoid displacement, which can otherwise hinder sufficiently deep deposition and thereby prevent the radiative collapse and subsequent transport event. In the hot-background limit, where Alfv\'enic currents inside the plasmoid are negligible, the drift distance scales approximately as $\Delta r \propto B^{-2}$ so that the factor-two lower field in DD7.5MA compared to the $15\,\mathrm{MA}$ DTHmode24 scenario implies a $\sim$fourfold larger drift (cf.\ Eq.~(A6) in \citep{Vallhagen2023}). Consistent with this scaling, simulations of DD7.5MA with an artifical $B\simeq 5.3\,\mathrm{T}$ magnetic field (i.e. comparable to the $15\,\mathrm{MA}$ scenarios) exhibit plasmoid drifts similar to DTHmode24. Moreover, in this higher-$B$ setting, simulations using a substantially reduced neon fraction (a factor $\sim 5$ lower than the baseline DD7.5MA mixture) yield negligible representative runaway current even for a single injection. These additional tests indicate that the apparent ``narrow'' mitigation window in the baseline $7.5\,\mathrm{MA}$ study is, to some extent, contingent on the assumed drift amplitude; improving constraints on the scaling of $\Delta y$ in the model with $B$ and background parameters will be crucial for robust predictions.

\section{Discussion and conclusions}
\label{sec:discussion}

\begin{figure*}
    \centering
    \begin{overpic}[width=1\linewidth]{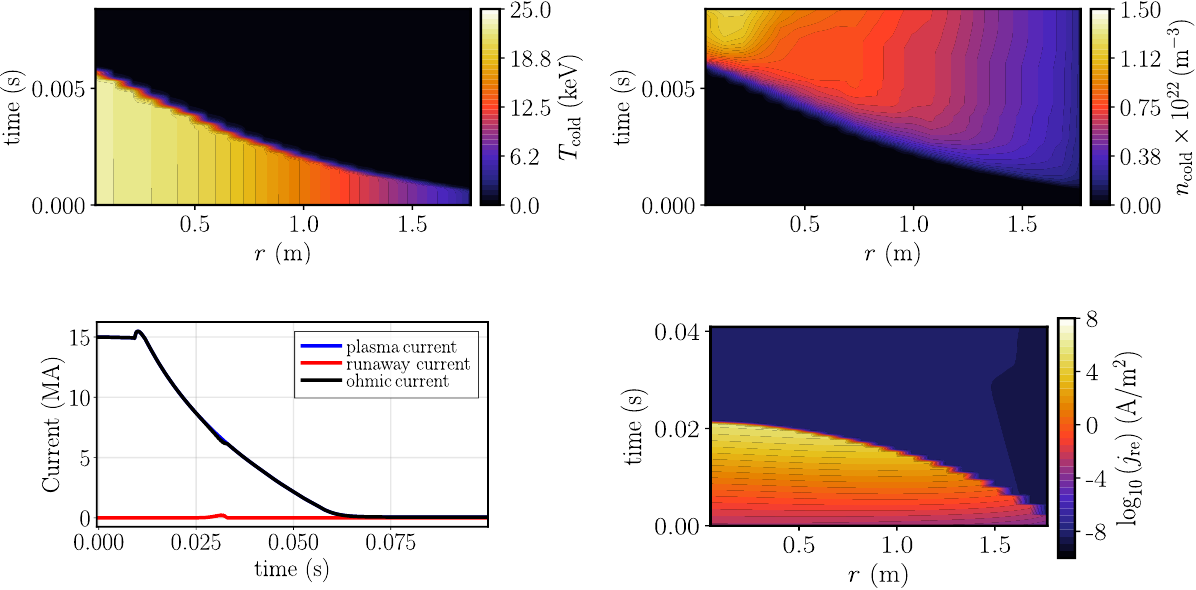}
        \put(7.5,49.5){\textbf{(a)}}
        \put(59,49.5){\textbf{(b)}}
        \put(7.5,23.5){\textbf{(c)}}
        \put(59,23){\textbf{(d)}}
    \end{overpic}
    \caption{Two--stage SPI strategy for ITER that maximizes assimilation while minimizing runaway retention. Stage~1 uses an almost pure D pellet with a minimal Ne doping (\(\sim0.1\%\)) which is assumed sufficient to suppress the plasmoid drifts, improving core deposition \emph{without} triggering a premature TQ (thus avoiding an early cut-off of assimilation). 
Stage~2 follows after 5 ms with a Ne-rich pellet (\(f_{\mathrm{Ne}}\approx5\%\)) to induce a controlled radiative TQ after the plasma has been diluted and cooled, suppressing hot-tail production and limiting avalanche growth. 
Panels show: (a) spatio-temporal evolution of the electron temperature \(T_\mathrm{cold}(r,t)\) prior to the TQ; 
(b) corresponding electron density \(n_\mathrm{cold}(r,t)\); 
(c) time traces of total plasma current, ohmic current, and runaway current from injection to termination; 
(d) logarithm of the runaway current density \(j_{\mathrm{re}}(r,t)\) during the CQ, highlighting near-edge depletion consistent with scrape-off losses.}

    \label{fig:ITERbest}
\end{figure*}
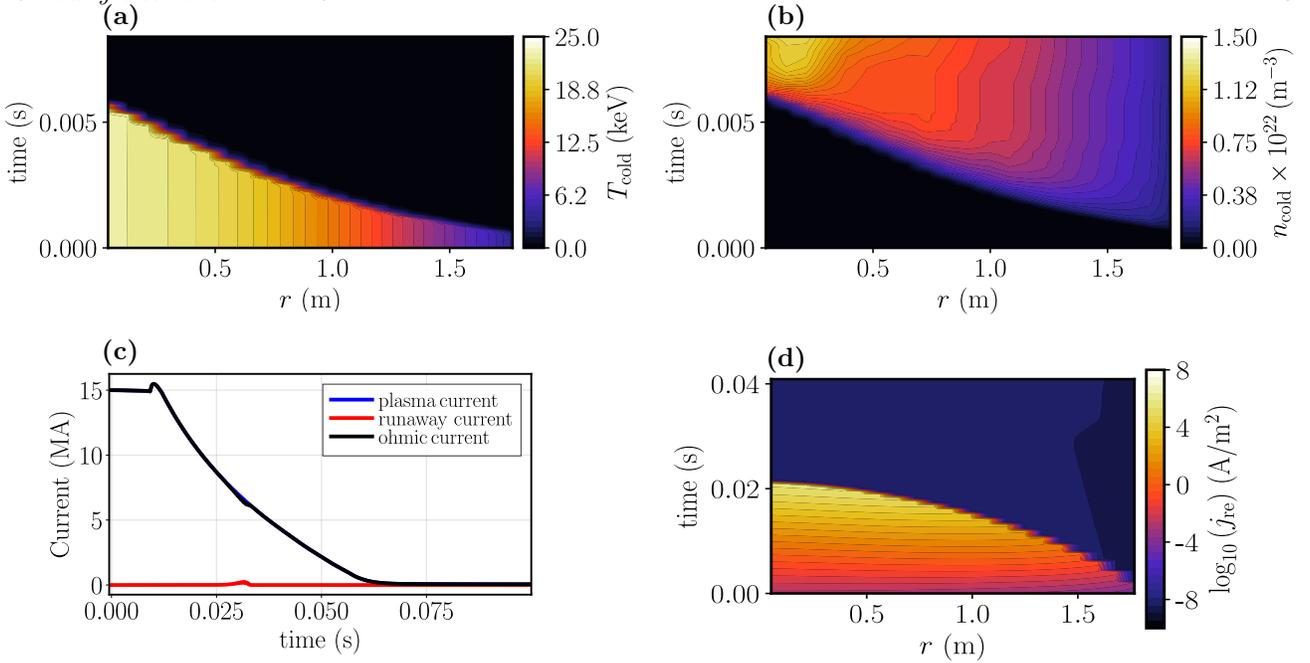

The present study provides a comprehensive overview of simulations carried out using the 1D code \DREAM\ of the RE dynamics in a wide range of SPI-mitigated ITER disruption scenarios, revealing a consistent picture across the parameter scan: it could be possible to limit RE generation to acceptable levels in both non-nuclear and nuclear scenarios in ITER, but in a confined range of its operational parameter space. The key findings can be summarized as:

\begin{enumerate}
        \item \textbf{Plasmoid drifts make SPI challenging in reactor-grade plasmas.}
    The outcomes of mitigation strategies, particularly staggered injections, differ between L-mode and H-mode plasmas. H-mode plasmas, such as the DD7.5MA and DTHmode24 scenarios, exhibit an \emph{H-mode penalty}. This is due to the fact that, in H-mode, pellets tend to ablate closer to the plasma edge, so the ablation cloud is formed at larger radii. As a result, the same plasmoid drift displaces the material more effectively towards the low-field side, hindering core deposition. This can lead to hollow density profiles, and a failure to suppress the hot-tail. In contrast, L-mode plasmas which have lower initial thermal energy content experience weaker internal pressure within the ablation cloud, resulting in slower drift velocities and thus less outward displacement of ablated material. Consequently, staggered injections tend to perform better in L-mode  but can fail in a hot H-mode plasma unless additional measures are taken to suppress the drift, e.g.\ by doping the first pellet with a small amount of neon \cite{Vallhagen2025b}. Note, however, that experimentally it has been found that even trace neon can enhance radiative cooling and may trigger a prompt TQ, as reported for JET trace-Ne SPI discharges \cite{JachmichFEC2025,KongFEC2025}.
        \item\textbf{VDEs effectively suppress REs in near-mitigated scenarios.}
    Runaway scrape-off losses associated with VDEs can play a decisive role in determining whether a disruption produces a residual RE current or a multi-MA beam. During a VDE, once the LCFS is lost and field lines open, a significant fraction of the plasma current is converted into halo currents. These halo currents can retain part of the poloidal flux, thereby reducing the energy available for the avalanche multiplication. This effect reduces the effective avalanche gain and makes seed suppression more efficient.
    The effectiveness of the scrape-off effect is however strongly scenario dependent and highly sensitive to the current-profile evolution during the TQ. Since the VDE and corresponding opening of flux surfaces is itself driven by the flux consumption (or current decay) in this cold VDE limit, this effect is successful in preventing REs only when the complete opening of field lines precedes the generation of macroscopic RE currents. 
        \item\textbf{Compton seed can be dominant in nuclear scenarios.}
    Across all injection schemes, nuclear seed production, in particular Compton electrons driven by energetic \(\gamma\)-rays, can dominate the runaway source. In cases with poor mitigation, however, the hot-tail seed and (where relevant) the tritium $\beta$-decay seed may be comparable to or exceed the Compton seed. We emphasize that even when the Compton mechanism dominates the \textit{seed}, it is not necessarily the dominant contributor to the final runaway current, which is typically set by avalanche multiplication acting on the total seed. This channel carries substantial uncertainty because the \(\gamma\)-ray flux from the ITER first wall during disruptions is not experimentally proven and must be inferred from modelling. In the nuclear scenarios considered here, we adopt the conservative assumption that nuclear reactions promptly cease when the TQ completes, and reduce in intensity by a factor $10^{-4}$. In practice, well-confined, fusion-producing plasmas are unlikely to be the target of mitigation, so that nuclear reactions may diminish even before pellet injection, thus implying that our simulations likely overestimate the nuclear seed sources. Nevertheless, once a macroscopic runaway current is generated and avalanche multiplication dominates, the representative runaway current depends only weakly on the seed amplitude. Consequently, uncertainties in the nuclear seed must be very large (i.e. orders of magnitude) to produce a sizeable change in the representative RE current. Nonetheless, when nuclear seed sources are artificially disabled in the model, DT H-mode cases with multiple pellets reach negligible representative runaway current under favourable (long) TQ conditions.

        \item\textbf{Low-Ne injections can be effective in suppressing REs in non-nuclear plasmas.}
    The analysis shows that reducing the neon fraction in injected pellets systematically decreases the representative runaway current in L-mode non nuclear plasmas, primarily by extending the pre-TQ phase and allowing the suprathermal electron tail to thermalize. This suppresses the hot-tail seed and limits the avalanche gain by reducing the total number of bound electrons in the plasma. However, in H-mode DT nuclear plasmas, this benefit is offset by the dominant Compton-driven nuclear seed, which remains largely unaffected by $f_{\rm Ne}$, consistently producing a multi-MA runaway beam under ITER-relevant conditions with our baseline model. As a result, while lowering $f_{\rm Ne}$ can effectively mitigate runaways in L-mode scenarios, it is insufficient on its own to suppress runaway current in DT H-mode plasmas, where the Compton seed sets a strict floor for mitigation performance.

        \item\textbf{Current density profile relaxation during TQ can limit avalanche gain.}
    By including a hyper-resistive term to mimic MHD-induced current flattening, we observed that the avalanche gain is effectively reduced beyond what only the VDE scrape-off model achieves. The flatter current profile and reduced poloidal-flux variation provides a potential reduction of the avalanche multiplication by 4 orders of magnitude in the cases considered. While this reduction can substantially decrease the final runaway current, it is not, by itself, sufficient to guarantee RE avoidance in ITER-relevant scenarios where the unmitigated avalanche multiplication can be extremely large.        
        \item\textbf{CQ radial transport can greatly reduce the RE current.}
    The assumption of perfect RE confinement inside the LCFS during CQ likely underestimates radial losses. Strong MHD activity can lead to stochastic magnetic fields and enhanced radial transport of runaways. Introducing a Rechester–Rosenbluth-type diffusive transport of REs during the CQ shows a clear threshold behavior in the representative RE current. For weak magnetic perturbations $\delta B/B \leq 10^{-4}$, multi-MA runaway currents develops. However, once the perturbation strength exceeds a critical value, around a few $10^{-4}$, the final runaway current drops to negligible levels. For $\delta B/B \approx 4 \times 10^{-4}$, the RE current was suppressed in all cases studied. This suggests that strong magnetic perturbations (e.g.\ external 3D fields, as studied in~\cite{Yoshino2000,Lehnen2008,Papp2011}) could be a complementary strategy to reduce avalanche growth. While previous studies of resonant magnetic perturbations in ITER found little effect on core runaways, we suggest that the perturbations may prove more effective when the vertical plasma motion is also accounted for.

        \item\textbf{RE avoidance possible also in reduced-current scenario.}
    Encouragingly, simulations indicate that runaway avoidance is achievable at lower plasma currents, which is relevant for the early phases of operation in ITER. In a \qty{7.5}{MA} H-mode scenario, complete RE suppression was achieved with a staggered injection assuming a long TQ.
        The \(7.5\,\mathrm{MA}\) case exhibits the same drift-induced H-mode assimilation issues as the full current scenarios, but the runaway conversion efficiency is inherently limited at reduced current because of the smaller avalanche gain. This allows a much larger initial seed of up to \(\sim 10^{-1}\,\mathrm{A}\) to be tolerated in the simulations thanks to the VDE scrape-off mechanism. In practice, mitigation in the \qty{7.5}{MA} plasma was achieved by doping the first protium pellet with a small ($0.1\%$) amount of neon, which enhances radiation and slows the plasmoid drift, thereby improving assimilation while still maintaining an extended pre-TQ phase. As a result, even with realistic seeds, the runaways were dissipated and the total RE current remained below concerning levels.
\end{enumerate}

Building on these insights, figure~\ref{fig:ITERbest} showcases a proposed two–stage SPI disruption mitigation scheme in full-current ITER H-mode DT plasmas and highlights why it is so effective at suppressing runaway electrons. In this case, the first injection is composed of three protium pellets with a small admixture of neon. The protium payload is chosen to raise $n_e$ and mainly to thermalize RE seeds and also reduce the avalanche gain, while avoiding excessive fueling that would trigger recombination and ultimately reduce \(n_e\). A minimal $0.1\%$ neon fraction is assumed to be sufficient to cool the ablation cloud and limit its drift, thereby improving penetration and core deposition. This small impurity content does not produce sufficient radiative cooling to trigger a premature TQ, ensuring that material assimilation is not interrupted too early.

A Ne-rich pellet ($f_\mathrm{Ne} = 5\%$) follows in the second stage in order to induce a TQ in \qty{8.4}{ms} once the plasma is already diluted and cooled by the first stage, consequently suppressing the hot-tail generation efficiently. In this case, a substantial amount of protium is assimilated
while only a modest neon quantity
assimilates. This mitigation scheme produces a representative runaway seed of \qty{15}{mA}, consistent with the most favorable DT H-mode ITER scenarios, where hot-tail generation is effectively suppressed and the seed population is instead dominated by Compton scattering. However, this simulation presents a viable theoretical pathway to limit the runaway electron beam even in presence of nuclear seeds. In fact, in this scenario, even if the RE seed is large compared to non-nuclear cases, it only yields a representative RE current of \qty{0.24}{MA}. This is because of the inclusion of hyper-resistive diffusion in the TQ and the VDE scrape-off model, which leads to a broader post-TQ current profile and significant reduction of the avalanche gain. This synergy allows ITER to tolerate a much larger runaway seed than previously anticipated, up to mA level in our sensitivity scans, without a dramatic runaway avalanche multiplication. However, this outcome relies on the assumed degree of current-profile flattening during the TQ and on the post-TQ evolution, which cannot be controlled a priori and may be less favorable in ITER.

Including hyper-resistive diffusion during the TQ broadens the post-TQ current profile and induces a corresponding reduction of the poloidal flux, in turn reducing the available avalanche gain. In such scenario, even when the RE seed is large compared to non-nuclear cases, the final RE current remains limited, reaching a representative value of $0.24\,\mathrm{MA}$. This simulation therefore illustrates a viable theoretical pathway to limit the RE beam even in the presence of nuclear seeds: the synergy between TQ current profile flattening and vertical losses allows ITER to tolerate substantially larger seeds (up to the mA level in our sensitivity scans) without triggering dramatic avalanche multiplication. At the same time, the broader set of simulations indicates that robust RE avoidance in ITER remains challenging and appears attainable only within a limited set of conditions that are not fully controllable (e.g.\ strong TQ flattening and/or enhanced CQ transport). In this sense, ITER lies close to a threshold where RE avoidance is only marginally accessible, and extrapolation to higher-current devices may be less favourable due to the larger intrinsic avalanche gain, in the absence of additional loss channels such as transport driven by three-dimensional fields.

\section*{Acknowledgement}
The authors are grateful to C Wang, S Jachmich, I Ekmark and P Halldestam for fruitful discussion. This work has been carried out in collaboration with the ITER Organisation under an implementing agreement co-funded by the ITER DMS Task Force. ITER is the Nuclear Facility INB No.\ 174. This paper explores physics processes during the plasma operation of the tokamak when disruptions take place; nevertheless, the nuclear operator is not constrained by the results presented here. The views and opinions expressed herein do not necessarily reflect those of the ITER Organization. This work has been partially carried out within the framework of the EUROfusion Consortium, funded by the European Union via the Euratom Research and Training Programme (Grant Agreement No 101052200 — EUROfusion). Views and opinions expressed are however those of the author(s) only and do not necessarily reflect those of the European Union or the European Commission. Neither the European Union nor the European Commission can be held responsible for them. This work was supported by the Swedish Research Council (Dnr.\ 2024-04879).
\bibliography{biblio.bib}

\appendix
\section{Summary of simulation parameters}
This appendix provides lists of the main parameters for the simulations conducted in this paper. Table~\ref{tab:dBB} provides the magnetic perturbation strength $\delta B/B$ of the TQ used in the three different ITER scenarios, at the two different TQ durations used. As described in section~\ref{sec:theory:disruption}, the magnetic perturbation is set as to ensure that the temperature drops below \qty{200}{eV} during the TQ.

Table~\ref{tab:scenarios} lists the main parameters used in the basline SPI cases studied. These parameters are identical to those used for the same cases in Ref.~\cite{Vallhagen2024}.

\begin{table}[h!]
    \centering
    \caption{Magnetic perturbation amplitude $\delta B/B$ calculated to give TQ times of 1 or 3 ms, for the four scenarios studied in this document.}
    \label{tab:dBB}
    \begin{tabular}{lcc}
        \hline
        Scenario & $t_{\mathrm{TQ}}$ (ms) & $\delta B / B$ (\%) \\\hline
        H26         & 1 & 0.359 \\
        H26         & 3 & 0.207 \\
        \hline
        DTHmode24   & 1 & 0.320 \\
        DTHmode24   & 3 & 0.185 \\
        \hline
        DD7.5       & 1 & 0.333 \\
        DD7.5       & 3 & 0.192 \\
        \hline
    \end{tabular}
\end{table}

\begin{table*}
\centering

\caption{Overview of all simulated baseline scenarios, including main (S), multi-pellet (M), and staggered (St) cases with corresponding TQ timing, neon content, and pellet number. For the staggered cases, the pellet number refers to the number of pure hydrogenic pellets injected during the initial injection. The total number of pellet shards is fixed at the default value of 487 fragments, except for cases S17 and S18, where it is set to 68 and 5185 fragments, respectively. The corresponding Hydrogenic atoms per pellet can be calculated as $185 \times10^{22}$ atoms minus the corresponding Ne atoms per pellet}
\label{tab:scenarios}
\begin{tabular}{cllllll}
\toprule
\textbf{No} & \textbf{Target plasma} & $t_{\rm TQ}$ & \textbf{TQ onset} & \textbf{Pellets} & \textbf{Ne atoms per pellet [$10^{22}$]} \\
\midrule
S1  & H26         & \qty{3}{ms} & Late  &   1   & 183 & \\
S2  & H26         & \qty{3}{ms} & Late  &   1   & 20 & \\
S3  & H26         & \qty{3}{ms} & Late  &   1   & 5.22 & \\
S4  & H26         & \qty{1}{ms} & Early &1& Same as S1 & \\
S5  & H26         & \qty{1}{ms} & Early & 1     & Same as S2 & \\
S6  & H26         & \qty{1}{ms} & Early &   1   & Same as S3 & \\
S7  & H26         & \qty{1}{ms} & Late  &  1    & Same as S1 & \\
S8  & H26         & \qty{3}{ms} & Early &   1   & Same as S1 & \\
S9  & H26         & \qty{3}{ms} & Late  &  1    & Same as S2 & \\
S10 & H26         & \qty{3}{ms} & Late  &   1   & Same as S2 & \\
S11 & DTHmode24   & \qty{3}{ms} & Late  &   1   & 150 & \\
S12 & DTHmode24   & \qty{3}{ms} & Late  &   1   & 2.5 & \\
S13 & DTHmode24   & \qty{3}{ms} & Late  &  1    & 5.22 & \\
S14 & DTHmode24   & \qty{1}{ms} & Early &   1   & Same as S11 & \\
S15 & DTHmode24   & \qty{1}{ms} & Early &   1   & Same as S12 & \\
S16 & DTHmode24   & \qty{1}{ms} & Early &  1    & Same as S12 & \\
M1  & H26         & \qty{3}{ms} & Late  & 2    & 1/2 of S2 & \\
M2  & H26         & \qty{3}{ms} & Late  & 3    & 1/3 of S2 & \\
M3  & H26         & \qty{1}{ms} & Early & 2    & 1/2 of S2 & \\
M4  & H26         & \qty{1}{ms} & Early & 3    & 1/3 of S2 & \\
M5  & DTHmode24   & \qty{3}{ms} & Late  & 2    & 1/2 of S12 & \\
M6  & DTHmode24   & \qty{3}{ms} & Late  & 3    & 1/3 of S12 & \\
M7  & DTHmode24   & \qty{1}{ms} & Early & 2    & 1/2 of S12 & \\
M8  & DTHmode24   & \qty{1}{ms} & Early & 3    & 1/3 of S12 & \\
St1 & H26         & \qty{3}{ms}     &  Late     & 1    & Same as S2  & \\
St2 & H26         & \qty{3}{ms}     &  Late     & 2    & Same as S2  & \\
St3 & H26         & \qty{3}{ms}     &  Late     & 3    & Same as S2  & \\
St4 & DTHmode24   & \qty{3}{ms}     &  Late     & 1    & Same as S12 & \\
St5 & DTHmode24   & \qty{3}{ms}     &  Late     & 2    & Same as S12 & \\
St6 & DTHmode24   & \qty{3}{ms}     & Late      & 3    & Same as S12 & \\
\bottomrule
\end{tabular}
\end{table*}

\end{document}